\documentclass[preprint,aps,showpacs,showkeys]{revtex4}
\usepackage{supertabular}
\usepackage{amsfonts}
\usepackage{amsmath}%fig
\usepackage{amsfonts}%fig
\usepackage{amssymb} %fig
\usepackage{amsxtra} %fig
\begin{document}
\title{The {\it spin-charge-family} theory offers understanding of the triangle anomalies 
cancellation in the {\it standard model}  
 }
\author{N.S. Manko\v c Bor\v stnik}
 \affiliation{University of Ljubljana}
\email{norma.mankoc@fmf.uni-lj.si}
%Slovenia
\author{H.B.F. Nielsen}
\affiliation{Niels Bohr Institute} 
\email{hbech@nbi.dk}
%Copenhagen, DK-2100 
%}
% 

\begin{abstract}
The {\it standard model} has for massless quarks and leptons "miraculously" no triangle anomalies 
due to the fact that the sum of all possible traces $Tr [\tau^{Ai}$$ \tau^{Bj}$$ \tau^{Ck}]$ ---
where $\tau^{Ai}, \tau^{Bi}$ and $\tau^{Ck}$ are the generators of one, of two or of three of 
the groups $SU(3), SU(2)$ and $U(1)$ --- over the representations of one family of the left handed 
fermions and anti-fermions (and separately of the right handed fermions and anti-fermions), 
contributing to the triangle currents, is equal to 
zero~\cite{Alvarez,AlvarezWitten,Bilal,AlvarezBondiaMartin}.  
It is demonstrated in this paper that this cancellation of the {\it standard model} triangle anomaly 
follows straightforwardly if the $SO(3,1), SU(2), U(1)$ and $ SU(3)$  are the subgroups of the 
orthogonal group $SO(13,1)$, as it is in the {\it spin-charge-family} theory~\cite{EPJC2017,JMP2015,%
norma2014MatterAntimatter,NBled2013,NBled2012,IARD2016,pikanorma,portoroz03,norma92,%
norma93,norma94,norma95,gmdn07,gn,gn2013,gn2015,NPLB,N2014scalarprop}. We comment 
on the $SO(10)$ anomaly cancellation, which works if handedness and charges are related
 "by hand".

\end{abstract}


\maketitle

 \section{Introduction} 
 \label{introduction}

In $d=(2n)$-dimensional space-time massless fermions contribute through the one-loop (n + 1)-angle 
diagram in general an anomalous (infinite) function, which causes the current non-conservation and 
contributes to the gauge non-invariance of the 
action~\cite{Alvarez,AlvarezWitten,Bilal,AlvarezBondiaMartin}.

We discuss anomalies in $[d=(3+1)]$-dimensional space-time, that is the triangle anomalies cancellation, 
from the point of view of the {\it spin-charge-family} theory~\cite{EPJC2017,JMP2015,%
norma2014MatterAntimatter,%
NBled2013,NBled2012,IARD2016,pikanorma,portoroz03,norma92,norma93,norma94,norma95,%
gmdn07,gn,gn2013,gn2015,NPLB,N2014scalarprop} (which starts with a simple action in 
$d=(13+1)-$dimensional space-time~\cite{JMP2015,IARD2016}%
) to demonstrate that embedding the {\it standard model} 
groups into the orthogonal group $SO(13,1)$ explains elegantly the "miraculous" cancellation of 
the triangle anomalies in the {\it standard model}, explaining at the same time also all the 
assumptions of the {\it standard model} concerning the fermion charges and their connection 
with the handedness and the properties of their corresponding anti-fermions. 

We comment also on the $SO(10)$ explanation of the anomaly cancellation from the point of 
view of the {\it spin-charge-family} theory: While one fundamental representation of $SO(13,1)$  
contains the left handed members of one family with the weak and hyper charges as required by
 the {\it standard model} and the right handed family members, which are weak chargeless  
with the hyper charge as required by the {\it standard model}, as well as the corresponding
right handed weak charged and left handed weak chargeless antiparticles, 
%~\footnote{The antiparticle states of quarks and leptons, both appearing in the same representation 
%presented in Table~\ref{Table so13+1.}, can be reached also by applying on the particle states the 
%operator $\mathbb{C}_{{\cal N}}  {\cal P}_{{\cal N}}$~\cite{discretesym}.},  
the $SO(10)$ unification scheme must relate  the spin and handedness with the charges "by hand". 
%(like the {\it standard model} does) to explain  the "miracle" of the {\it standard model} 
%cancellation of the triangle anomaly. 
The  {\it spin-charge-family} theory explains, by using our technique~\cite{norma93,hn02}, 
the $SO(10)$ unification scheme~\footnote{The {\it spin-charge-family} theory explains 
also, what the $SO(10)$ unifying theory can not without adding "by hand" new groups: The origin 
of families, the origin of matter-antimatter asymmetry, the appearance of the dark matter and 
other phenomena~\cite{IARD2016,JMP2015,norma2014MatterAntimatter}.} transparently and in another (nonstandard and correspondingly an useful) way.

To the  triangle anomaly the right-handed spinors (fermions) and the right handed anti-spinors
(anti-fermions) contribute with the opposite sign than the left handed spinors and the left handed 
anti-spinors. Their common contribution to anomalies is proportional to~\cite{Bilal} 
\begin{eqnarray}
\label{anomaly0}
(\sum_{(A,i,B,j,C,k)_{L \,\bar{L}}} Tr [\tau^{Ai} \, \tau^{Bj} \,\tau^{Ck}] -
 \sum_{(A,i,B,j,C,k)_{R \,\bar{R}}} Tr [\tau^{Ai} \, \tau^{Bj} \,\tau^{Ck}]\,) \,,
\end{eqnarray}
where $\tau^{Ai}$ are in the {\it standard model} the generators of the infinitesimal transformation 
of the groups $SU(3), SU(2)$ and $U(1)$, while in the {\it spin-charge-family} theory
$\tau^{Ai}$ are the infinitesimal generators of the irreducible subgroups of the starting orthogonal 
group $SO(2(2n+1)-1,1)$, $n=3$ (which include all the {\it standard model} groups, offering
correspondingly the explanation for their origin). 
%The sign ${}_{L \,\bar{L}}$  (${}_{R \,\bar{R}}$) denotes the left (right) handed  spinors  and 
%their anti-spinors (right (left)), respectively. 
The traces run over the representations of one massless family of the left handed fermions and 
anti-fermions, denoted by ${}_{L \,\bar{L}}$, and the right handed fermions and anti-fermions, denoted 
by ${}_{R \,\bar{R}}$.
 
In $d=(2n)$-dimensional space-time the summation runs over $(n+1)$ products of generators.

One Weyl representation of spinors (the fundamental representation with $2^{\frac{d}{2}-1}$
states, $64$ in $d=13+1$) of $SO(13,1)$ contains, if analyzed from the point 
of view of the {\it standard model} groups, all the left and the right handed quarks and leptons, and 
the right handed anti-partners of the left handed quarks and leptons and the left handed anti-partners 
of the right handed quarks and leptons of one family, explaining the assumed properties of the 
{\it standard model}. Each state of spinors appear in this fundamental representation of 
$SO(13,1)$ with the spin up and with the spin down. 
The application of the operator $\mathbb{C}_{{\cal N}}  {\cal P}_{{\cal N}} $ on any quark or 
lepton state 
transforms this state into its antiparticle state of the same spin and the opposite handedness, 
belonging to the same representation~\footnote{Discrete symmetries in $d=(3+1)$ follow from the 
corresponding definition of these symmetries in $d$-dimensional space~\cite{discretesym,TDN}.}. 

All these states are presented in Table~\ref{Table so13+1.}  in the "technique" with nilpotents and 
projectors~\cite{hn02}. 
Table~\ref{Table so13+1.}  presents spinor handedness ($\Gamma^{(3,1)}$), their spin ($S^{12}$),
weak charge $(\tau^{13})$, the second $SU(2)_{II}$ $(\tau^{23})$ charge (arising together with
$SU(2)_{I}$ from $SO(4)$), their colour charge $(\tau^{33},\tau^{38})$  (arising together with
$U(1)_{II}$ from $SO(6)$), and the "fermion charge" ($\tau^{4}$, the generator of $U(1)_{II}$). 
The hyper charge is $Y=(\tau^{23} +\tau^{4})$,  the electromagnetic charge is 
$Q=(\tau^{13}+ Y)$. All these operators  ($\tau^{Ai}$) are defined in Sect.~\ref{SCFT} 
(Eqs.~(\ref{so42},\ref{so64})).
The reader is kindly advised to follow the explanation in Subsect.~\ref{explicit}, in particular 
{\it Example 1.} and {\it Example 2.},  to learn how can one read properties of each state from 
niloptents and projectors, the 
product of which forms in the "technique", Subsect.~\ref{technique}, each state of the fundamental
 representation.

To make it easier for the reader to compare  the states of one Weyl representation of $SO(13,1)$ 
with the by the {\it standard model} assumed presentation of one family of quarks and leptons,
let us point out that the {\it standard model} content presents separately states and anti-states,
does not pay attention on spin up and down and does not include the right handed neutrino as the 
regular member. The one Weyl representation of $SO(13,1)$  includes $64$ states:
($u_{L}^{ci}, d_{L}^{ci},  u_{R}^{ci}, d_{R}^{ci}$), $i=1,2,3$ and  ($\nu_{L}, e_{L}, \nu_{R}, 
e_{R}$), all with spin up and down, representing $32$ states, the states of the corresponding 
anti-quarks and anti-leptons with spin up and down contribute another $32$ states.

Let it be commented also that the $SO(10)$ content of states (we shall comment these unification 
theories from the point of view of one Weyl representation of $SO(13,1)$ in Sect.~\ref{so10} and in 
Subsect.~\ref{explicit},  {\it Example 4.}) has in one fundamental
representation of $SO(10)$ $2^{\frac{10}{2}-1} =16$ states with charges and anti-charges,  while 
the  relation between charges and handedness, left or right, and spin (up and down) which appear 
in $SO(3,1)$, must be added  "by hand". With the handedness and spins added the enlarged 
$SO(10)$ has again $64$ states. 
($SO(10)$ unifying theories offer no explanation for families and for the  
origin of the vector gauge fields and of the scalar fields.)

%%%%%%%%%%%%%%%%

%\begin{tiny} 
\bottomcaption{\label{Table so13+1.}%
\tiny{
The left handed ($\Gamma^{(13,1)} = -1$, 
% ($ = \Gamma^{(7,1)} \times \Gamma^{(6)}$),  
Eq.~(\ref{hand})) 
multiplet of spinors --- the members of the fundamental representation of the $SO(13,1)$ group, 
manifesting the subgroup $SO(7,1)$ 
 of the colour charged quarks and anti-quarks and the colourless 
leptons and anti-leptons --- is presented in the massless basis using the technique presented in
App.~\ref{technique}. %~\cite{snmb:hn02hn03}. 
It contains the left handed  ($\Gamma^{(3,1)}=-1$) weak ($SU(2)_{I}$) charged  
($\tau^{13}=\pm \frac{1}{2}$, Eq.~(\ref{so42})), 
%$\vec{\tau}^{1}= \frac{1}{2} (S^{58}- S^{67}, S^{57}+ S^{68}, S^{56}- S^{78})$) 
and $SU(2)_{II}$ chargeless ($\tau^{23}=0$, Eq.~(\ref{so42}))
%$\vec{\tau}^{2}= \frac{1}{2} (S^{58}+ S^{67}, S^{57}- S^{68}, S^{56}+ S^{78})$)
quarks and leptons and the right handed  ($\Gamma^{(3,1)}=1$, Subsect.~\ref{explicit}) 
 weak  ($SU(2)_{I}$) chargeless and $SU(2)_{II}$
charged ($\tau^{23}=\pm \frac{1}{2}$) quarks and leptons, both with the spin $ S^{12}$  up and 
down ($\pm \frac{1}{2}$, respectively). 
Quarks distinguish from leptons only in the $SU(3) \times U(1)$ part: Quarks are triplets 
of three colours  ($c^i$ $= (\tau^{33}, \tau^{38})$ $ = [(\frac{1}{2},\frac{1}{2\sqrt{3}}), 
(-\frac{1}{2},\frac{1}{2\sqrt{3}}), (0,-\frac{1}{\sqrt{3}}) $], Eq.~(\ref{so64}))
%$\vec{\tau}^{3}= \frac{1}{2}(S^{9\,12}- S^{10\,11},S^{9\,11}+ S^{10\,12},S^{9\,10}- 
%S^{11\,12},$ $S^{9\,14}- S^{10\,13},S^{9\,13}+ S^{10\,14},S^{11\,14}- S^{12\,13},$
%$S^{11\,13}+ S^{12\,14},\frac{1}{\sqrt{3}}(S^{9\,10}+ S^{11\,12} - 2S^{13\,14})$), 
carrying  the "fermion charge" ($\tau^{4}=\frac{1}{6}$, Eq.~(\ref{so64})).
%$=-\frac{1}{3}(S^{9\,10}+ S^{11\,12}+ S^{13\,14})$).
The colourless leptons carry the "fermion charge" ($\tau^{4}=-\frac{1}{2}$). 
The same multiplet contains also the left handed weak ($SU(2)_{I}$) chargeless and $SU(2)_{II}$ 
charged anti-quarks and anti-leptons and the right handed weak ($SU(2)_{I}$) charged and 
$SU(2)_{II}$ chargeless anti-quarks and anti-leptons. 
Anti-quarks distinguish from anti-leptons again only in the $SU(3) \times U(1)$ part: Anti-quarks are 
anti-triplets, % ($\bar{c}^i$ $= (\tau^{33}, \tau^{38})$), 
 carrying  the "fermion charge" ($\tau^{4}=-\frac{1}{6}$). 
%$=-\frac{1}{3}(S^{9\,10}+ S^{11\,12}+ S^{13\,14})$)
The anti-colourless anti-leptons carry the "fermion charge" ($\tau^{4}=\frac{1}{2}$). 
% $ S^{12}$ defines the ordinary spin %(which can also be read directly from the basic vector, both
%vectors  with both spins, $\pm \frac{1}{2}$. 
 $Y=(\tau^{23} + \tau^{4})$ is the hyper charge, the electromagnetic charge 
is $Q=(\tau^{13} + Y$).
The states of opposite charges (anti-particle states) are reachable  from the particle states (besides by
$S^{ab}$) also by the application of the discrete symmetry operator 
${\cal C}_{{\cal N}}$ ${\cal P}_{{\cal N}}$, presented in Refs.~\cite{discretesym,TDN} and in 
Subsect.~\ref{explicit}.
The vacuum state, 
%$|vac>_{fam}$, 
on which the nilpotents and projectors operate, is not shown. 
%The basis is the massless one. 
The reader can find this  Weyl representation also in
Refs.~\cite{norma2014MatterAntimatter,pikanorma,portoroz03,JMP2015} and in the references therein. 
}
}
\tablehead{\hline
i&$$&$|^a\psi_i>$&$\Gamma^{(3,1)}$&$ S^{12}$&
$\tau^{13}$&$\tau^{23}$&$\tau^{33}$&$\tau^{38}$&$\tau^{4}$&$Y$&$Q$\\
\hline
&& ${\rm (Anti)octet},\,\Gamma^{(7,1)} = (-1)\,1\,, \,\Gamma^{(6)} = (1)\,-1$&&&&&&&&& \\
&& ${\rm of \;(anti) quarks \;and \;(anti)leptons}$&&&&&&&&&\\
\hline\hline} 
\tabletail{\hline \multicolumn{12}{r}{\emph{Continued on next page}}\\}
\tablelasttail{\hline}
%\begin{table}
%\begin{small}
%\begin{tiny}
\begin{center}
\tiny{
\begin{supertabular}{|r|c||c||c|c||c|c||c|c|c||r|r|}
1&$ u_{R}^{c1}$&$ \stackrel{03}{(+i)}\,\stackrel{12}{(+)}|
\stackrel{56}{(+)}\,\stackrel{78}{(+)}
||\stackrel{9 \;10}{(+)}\;\;\stackrel{11\;12}{(-)}\;\;\stackrel{13\;14}{(-)} $ &1&$\frac{1}{2}$&0&
$\frac{1}{2}$&$\frac{1}{2}$&$\frac{1}{2\,\sqrt{3}}$&$\frac{1}{6}$&$\frac{2}{3}$&$\frac{2}{3}$\\
\hline 
2&$u_{R}^{c1}$&$\stackrel{03}{[-i]}\,\stackrel{12}{[-]}|\stackrel{56}{(+)}\,\stackrel{78}{(+)}
||\stackrel{9 \;10}{(+)}\;\;\stackrel{11\;12}{(-)}\;\;\stackrel{13\;14}{(-)}$&1&$-\frac{1}{2}$&0&
$\frac{1}{2}$&$\frac{1}{2}$&$\frac{1}{2\,\sqrt{3}}$&$\frac{1}{6}$&$\frac{2}{3}$&$\frac{2}{3}$\\
\hline
3&$d_{R}^{c1}$&$\stackrel{03}{(+i)}\,\stackrel{12}{(+)}|\stackrel{56}{[-]}\,\stackrel{78}{[-]}
||\stackrel{9 \;10}{(+)}\;\;\stackrel{11\;12}{(-)}\;\;\stackrel{13\;14}{(-)}$&1&$\frac{1}{2}$&0&
$-\frac{1}{2}$&$\frac{1}{2}$&$\frac{1}{2\,\sqrt{3}}$&$\frac{1}{6}$&$-\frac{1}{3}$&$-\frac{1}{3}$\\
\hline 
4&$ d_{R}^{c1} $&$\stackrel{03}{[-i]}\,\stackrel{12}{[-]}|
\stackrel{56}{[-]}\,\stackrel{78}{[-]}
||\stackrel{9 \;10}{(+)}\;\;\stackrel{11\;12}{(-)}\;\;\stackrel{13\;14}{(-)} $&1&$-\frac{1}{2}$&0&
$-\frac{1}{2}$&$\frac{1}{2}$&$\frac{1}{2\,\sqrt{3}}$&$\frac{1}{6}$&$-\frac{1}{3}$&$-\frac{1}{3}$\\
\hline
5&$d_{L}^{c1}$&$\stackrel{03}{[-i]}\,\stackrel{12}{(+)}|\stackrel{56}{[-]}\,\stackrel{78}{(+)}
||\stackrel{9 \;10}{(+)}\;\;\stackrel{11\;12}{(-)}\;\;\stackrel{13\;14}{(-)}$&-1&$\frac{1}{2}$&
$-\frac{1}{2}$&0&$\frac{1}{2}$&$\frac{1}{2\,\sqrt{3}}$&$\frac{1}{6}$&$\frac{1}{6}$&$-\frac{1}{3}$\\
\hline
6&$d_{L}^{c1} $&$\stackrel{03}{(+i)}\,\stackrel{12}{[-]}|\stackrel{56}{[-]}\,\stackrel{78}{(+)}
||\stackrel{9 \;10}{(+)}\;\;\stackrel{11\;12}{(-)}\;\;\stackrel{13\;14}{(-)} $&-1&$-\frac{1}{2}$&
$-\frac{1}{2}$&0&$\frac{1}{2}$&$\frac{1}{2\,\sqrt{3}}$&$\frac{1}{6}$&$\frac{1}{6}$&$-\frac{1}{3}$\\
\hline
7&$ u_{L}^{c1}$&$\stackrel{03}{[-i]}\,\stackrel{12}{(+)}|\stackrel{56}{(+)}\,\stackrel{78}{[-]}
||\stackrel{9 \;10}{(+)}\;\;\stackrel{11\;12}{(-)}\;\;\stackrel{13\;14}{(-)}$ &-1&$\frac{1}{2}$&
$\frac{1}{2}$&0 &$\frac{1}{2}$&$\frac{1}{2\,\sqrt{3}}$&$\frac{1}{6}$&$\frac{1}{6}$&$\frac{2}{3}$\\
\hline
8&$u_{L}^{c1}$&$\stackrel{03}{(+i)}\,\stackrel{12}{[-]}|\stackrel{56}{(+)}\,\stackrel{78}{[-]}
||\stackrel{9 \;10}{(+)}\;\;\stackrel{11\;12}{(-)}\;\;\stackrel{13\;14}{(-)}$&-1&$-\frac{1}{2}$&
$\frac{1}{2}$&0&$\frac{1}{2}$&$\frac{1}{2\,\sqrt{3}}$&$\frac{1}{6}$&$\frac{1}{6}$&$\frac{2}{3}$\\
\hline\hline
\shrinkheight{0.25\textheight}
9&$ u_{R}^{c2}$&$ \stackrel{03}{(+i)}\,\stackrel{12}{(+)}|
\stackrel{56}{(+)}\,\stackrel{78}{(+)}
||\stackrel{9 \;10}{[-]}\;\;\stackrel{11\;12}{[+]}\;\;\stackrel{13\;14}{(-)} $ &1&$\frac{1}{2}$&0&
$\frac{1}{2}$&$-\frac{1}{2}$&$\frac{1}{2\,\sqrt{3}}$&$\frac{1}{6}$&$\frac{2}{3}$&$\frac{2}{3}$\\
\hline 
10&$u_{R}^{c2}$&$\stackrel{03}{[-i]}\,\stackrel{12}{[-]}|\stackrel{56}{(+)}\,\stackrel{78}{(+)}
||\stackrel{9 \;10}{[-]}\;\;\stackrel{11\;12}{[+]}\;\;\stackrel{13\;14}{(-)}$&1&$-\frac{1}{2}$&0&
$\frac{1}{2}$&$-\frac{1}{2}$&$\frac{1}{2\,\sqrt{3}}$&$\frac{1}{6}$&$\frac{2}{3}$&$\frac{2}{3}$\\
\hline
%$\cdots$&&&&&&&&&&&\\
%\hline 
11&$d_{R}^{c2}$&$\stackrel{03}{(+i)}\,\stackrel{12}{(+)}|\stackrel{56}{[-]}\,\stackrel{78}{[-]}
||\stackrel{9 \;10}{[-]}\;\;\stackrel{11\;12}{[+]}\;\;\stackrel{13\;14}{(-)}$
&1&$\frac{1}{2}$&0&
$-\frac{1}{2}$&$-\frac{1}{2}$&$\frac{1}{2\,\sqrt{3}}$&$\frac{1}{6}$&$-\frac{1}{3}$&$-\frac{1}{3}$\\
\hline 
12&$ d_{R}^{c2} $&$\stackrel{03}{[-i]}\,\stackrel{12}{[-]}|
\stackrel{56}{[-]}\,\stackrel{78}{[-]}
||\stackrel{9 \;10}{[-]}\;\;\stackrel{11\;12}{[+]}\;\;\stackrel{13\;14}{(-)} $
&1&$-\frac{1}{2}$&0&
$-\frac{1}{2}$&$-\frac{1}{2}$&$\frac{1}{2\,\sqrt{3}}$&$\frac{1}{6}$&$-\frac{1}{3}$&$-\frac{1}{3}$\\
\hline
13&$d_{L}^{c2}$&$\stackrel{03}{[-i]}\,\stackrel{12}{(+)}|\stackrel{56}{[-]}\,\stackrel{78}{(+)}
||\stackrel{9 \;10}{[-]}\;\;\stackrel{11\;12}{[+]}\;\;\stackrel{13\;14}{(-)}$
&-1&$\frac{1}{2}$&
$-\frac{1}{2}$&0&$-\frac{1}{2}$&$\frac{1}{2\,\sqrt{3}}$&$\frac{1}{6}$&$\frac{1}{6}$&$-\frac{1}{3}$\\
\hline
14&$d_{L}^{c2} $&$\stackrel{03}{(+i)}\,\stackrel{12}{[-]}|\stackrel{56}{[-]}\,\stackrel{78}{(+)}
||\stackrel{9 \;10}{[-]}\;\;\stackrel{11\;12}{[+]}\;\;\stackrel{13\;14}{(-)} $&-1&$-\frac{1}{2}$&
$-\frac{1}{2}$&0&$-\frac{1}{2}$&$\frac{1}{2\,\sqrt{3}}$&$\frac{1}{6}$&$\frac{1}{6}$&$-\frac{1}{3}$\\
\hline
15&$ u_{L}^{c2}$&$\stackrel{03}{[-i]}\,\stackrel{12}{(+)}|\stackrel{56}{(+)}\,\stackrel{78}{[-]}
||\stackrel{9 \;10}{[-]}\;\;\stackrel{11\;12}{[+]}\;\;\stackrel{13\;14}{(-)}$ &-1&$\frac{1}{2}$&
$\frac{1}{2}$&0 &$-\frac{1}{2}$&$\frac{1}{2\,\sqrt{3}}$&$\frac{1}{6}$&$\frac{1}{6}$&$\frac{2}{3}$\\
\hline
16&$u_{L}^{c2}$&$\stackrel{03}{(+i)}\,\stackrel{12}{[-]}|\stackrel{56}{(+)}\,\stackrel{78}{[-]}
||\stackrel{9 \;10}{[-]}\;\;\stackrel{11\;12}{[+]}\;\;\stackrel{13\;14}{(-)}$&-1&$-\frac{1}{2}$&
$\frac{1}{2}$&0&$-\frac{1}{2}$&$\frac{1}{2\,\sqrt{3}}$&$\frac{1}{6}$&$\frac{1}{6}$&$\frac{2}{3}$\\
\hline\hline
17&$ u_{R}^{c3}$&$ \stackrel{03}{(+i)}\,\stackrel{12}{(+)}|
\stackrel{56}{(+)}\,\stackrel{78}{(+)}
||\stackrel{9 \;10}{[-]}\;\;\stackrel{11\;12}{(-)}\;\;\stackrel{13\;14}{[+]} $ &1&$\frac{1}{2}$&0&
$\frac{1}{2}$&$0$&$-\frac{1}{\sqrt{3}}$&$\frac{1}{6}$&$\frac{2}{3}$&$\frac{2}{3}$\\
\hline 
18&$u_{R}^{c3}$&$\stackrel{03}{[-i]}\,\stackrel{12}{[-]}|\stackrel{56}{(+)}\,\stackrel{78}{(+)}
||\stackrel{9 \;10}{[-]}\;\;\stackrel{11\;12}{(-)}\;\;\stackrel{13\;14}{[+]}$&1&$-\frac{1}{2}$&0&
$\frac{1}{2}$&$0$&$-\frac{1}{\sqrt{3}}$&$\frac{1}{6}$&$\frac{2}{3}$&$\frac{2}{3}$\\
\hline
%$\cdots$&&&&&&&&&&\\
19&$d_{R}^{c3}$&$\stackrel{03}{(+i)}\,\stackrel{12}{(+)}|\stackrel{56}{[-]}\,\stackrel{78}{[-]}
||\stackrel{9 \;10}{[-]}\;\;\stackrel{11\;12}{(-)}\;\;\stackrel{13\;14}{[+]}$&1&$\frac{1}{2}$&0&
$-\frac{1}{2}$&$0$&$-\frac{1}{\sqrt{3}}$&$\frac{1}{6}$&$-\frac{1}{3}$&$-\frac{1}{3}$\\
\hline 
20&$ d_{R}^{c3} $&$\stackrel{03}{[-i]}\,\stackrel{12}{[-]}|
\stackrel{56}{[-]}\,\stackrel{78}{[-]}
||\stackrel{9 \;10}{[-]}\;\;\stackrel{11\;12}{(-)}\;\;\stackrel{13\;14}{[+]} $&1&$-\frac{1}{2}$&0&
$-\frac{1}{2}$&$0$&$-\frac{1}{\sqrt{3}}$&$\frac{1}{6}$&$-\frac{1}{3}$&$-\frac{1}{3}$\\
\hline
21&$d_{L}^{c3}$&$\stackrel{03}{[-i]}\,\stackrel{12}{(+)}|\stackrel{56}{[-]}\,\stackrel{78}{(+)}
||\stackrel{9 \;10}{[-]}\;\;\stackrel{11\;12}{(-)}\;\;\stackrel{13\;14}{[+]}$&-1&$\frac{1}{2}$&
$-\frac{1}{2}$&0&$0$&$-\frac{1}{\sqrt{3}}$&$\frac{1}{6}$&$\frac{1}{6}$&$-\frac{1}{3}$\\
\hline
22&$d_{L}^{c3} $&$\stackrel{03}{(+i)}\,\stackrel{12}{[-]}|\stackrel{56}{[-]}\,\stackrel{78}{(+)}
||\stackrel{9 \;10}{[-]}\;\;\stackrel{11\;12}{(-)}\;\;\stackrel{13\;14}{[+]} $&-1&$-\frac{1}{2}$&
$-\frac{1}{2}$&0&$0$&$-\frac{1}{\sqrt{3}}$&$\frac{1}{6}$&$\frac{1}{6}$&$-\frac{1}{3}$\\
\hline
23&$ u_{L}^{c3}$&$\stackrel{03}{[-i]}\,\stackrel{12}{(+)}|\stackrel{56}{(+)}\,\stackrel{78}{[-]}
||\stackrel{9 \;10}{[-]}\;\;\stackrel{11\;12}{(-)}\;\;\stackrel{13\;14}{[+]}$ &-1&$\frac{1}{2}$&
$\frac{1}{2}$&0 &$0$&$-\frac{1}{\sqrt{3}}$&$\frac{1}{6}$&$\frac{1}{6}$&$\frac{2}{3}$\\
\hline
24&$u_{L}^{c3}$&$\stackrel{03}{(+i)}\,\stackrel{12}{[-]}|\stackrel{56}{(+)}\,\stackrel{78}{[-]}
||\stackrel{9 \;10}{[-]}\;\;\stackrel{11\;12}{(-)}\;\;\stackrel{13\;14}{[+]}$&-1&$-\frac{1}{2}$&
$\frac{1}{2}$&0&$0$&$-\frac{1}{\sqrt{3}}$&$\frac{1}{6}$&$\frac{1}{6}$&$\frac{2}{3}$\\
\hline\hline
25&$ \nu_{R}$&$ \stackrel{03}{(+i)}\,\stackrel{12}{(+)}|
\stackrel{56}{(+)}\,\stackrel{78}{(+)}
||\stackrel{9 \;10}{(+)}\;\;\stackrel{11\;12}{[+]}\;\;\stackrel{13\;14}{[+]} $ &1&$\frac{1}{2}$&0&
$\frac{1}{2}$&$0$&$0$&$-\frac{1}{2}$&$0$&$0$\\
\hline 
26&$\nu_{R}$&$\stackrel{03}{[-i]}\,\stackrel{12}{[-]}|\stackrel{56}{(+)}\,\stackrel{78}{(+)}
||\stackrel{9 \;10}{(+)}\;\;\stackrel{11\;12}{[+]}\;\;\stackrel{13\;14}{[+]}$&1&$-\frac{1}{2}$&0&
$\frac{1}{2}$ &$0$&$0$&$-\frac{1}{2}$&$0$&$0$\\
\hline
27&$e_{R}$&$\stackrel{03}{(+i)}\,\stackrel{12}{(+)}|\stackrel{56}{[-]}\,\stackrel{78}{[-]}
||\stackrel{9 \;10}{(+)}\;\;\stackrel{11\;12}{[+]}\;\;\stackrel{13\;14}{[+]}$&1&$\frac{1}{2}$&0&
$-\frac{1}{2}$&$0$&$0$&$-\frac{1}{2}$&$-1$&$-1$\\
\hline 
28&$ e_{R} $&$\stackrel{03}{[-i]}\,\stackrel{12}{[-]}|
\stackrel{56}{[-]}\,\stackrel{78}{[-]}
||\stackrel{9 \;10}{(+)}\;\;\stackrel{11\;12}{[+]}\;\;\stackrel{13\;14}{[+]} $&1&$-\frac{1}{2}$&0&
$-\frac{1}{2}$&$0$&$0$&$-\frac{1}{2}$&$-1$&$-1$\\
\hline
29&$e_{L}$&$\stackrel{03}{[-i]}\,\stackrel{12}{(+)}|\stackrel{56}{[-]}\,\stackrel{78}{(+)}
||\stackrel{9 \;10}{(+)}\;\;\stackrel{11\;12}{[+]}\;\;\stackrel{13\;14}{[+]}$&-1&$\frac{1}{2}$&
$-\frac{1}{2}$&0&$0$&$0$&$-\frac{1}{2}$&$-\frac{1}{2}$&$-1$\\
\hline
30&$e_{L} $&$\stackrel{03}{(+i)}\,\stackrel{12}{[-]}|\stackrel{56}{[-]}\,\stackrel{78}{(+)}
||\stackrel{9 \;10}{(+)}\;\;\stackrel{11\;12}{[+]}\;\;\stackrel{13\;14}{[+]} $&-1&$-\frac{1}{2}$&
$-\frac{1}{2}$&0&$0$&$0$&$-\frac{1}{2}$&$-\frac{1}{2}$&$-1$\\
\hline
31&$ \nu_{L}$&$\stackrel{03}{[-i]}\,\stackrel{12}{(+)}|\stackrel{56}{(+)}\,\stackrel{78}{[-]}
||\stackrel{9 \;10}{(+)}\;\;\stackrel{11\;12}{[+]}\;\;\stackrel{13\;14}{[+]}$ &-1&$\frac{1}{2}$&
$\frac{1}{2}$&0 &$0$&$0$&$-\frac{1}{2}$&$-\frac{1}{2}$&$0$\\
\hline
32&$\nu_{L}$&$\stackrel{03}{(+i)}\,\stackrel{12}{[-]}|\stackrel{56}{(+)}\,\stackrel{78}{[-]}
||\stackrel{9 \;10}{(+)}\;\;\stackrel{11\;12}{[+]}\;\;\stackrel{13\;14}{[+]}$&-1&$-\frac{1}{2}$&
$\frac{1}{2}$&0&$0$&$0$&$-\frac{1}{2}$&$-\frac{1}{2}$&$0$\\
\hline\hline
33&$ \bar{d}_{L}^{\bar{c1}}$&$ \stackrel{03}{[-i]}\,\stackrel{12}{(+)}|
\stackrel{56}{(+)}\,\stackrel{78}{(+)}
||\stackrel{9 \;10}{[-]}\;\;\stackrel{11\;12}{[+]}\;\;\stackrel{13\;14}{[+]} $ &-1&$\frac{1}{2}$&0&
$\frac{1}{2}$&$-\frac{1}{2}$&$-\frac{1}{2\,\sqrt{3}}$&$-\frac{1}{6}$&$\frac{1}{3}$&$\frac{1}{3}$\\
\hline 
34&$\bar{d}_{L}^{\bar{c1}}$&$\stackrel{03}{(+i)}\,\stackrel{12}{[-]}|\stackrel{56}{(+)}\,\stackrel{78}{(+)}
||\stackrel{9 \;10}{[-]}\;\;\stackrel{11\;12}{[+]}\;\;\stackrel{13\;14}{[+]}$&-1&$-\frac{1}{2}$&0&
$\frac{1}{2}$&$-\frac{1}{2}$&$-\frac{1}{2\,\sqrt{3}}$&$-\frac{1}{6}$&$\frac{1}{3}$&$\frac{1}{3}$\\
\hline
35&$\bar{u}_{L}^{\bar{c1}}$&$\stackrel{03}{[-i]}\,\stackrel{12}{(+)}|\stackrel{56}{[-]}\,\stackrel{78}{[-]}
||\stackrel{9 \;10}{[-]}\;\;\stackrel{11\;12}{[+]}\;\;\stackrel{13\;14}{[+]}$&-1&$\frac{1}{2}$&0&
$-\frac{1}{2}$&$-\frac{1}{2}$&$-\frac{1}{2\,\sqrt{3}}$&$-\frac{1}{6}$&$-\frac{2}{3}$&$-\frac{2}{3}$\\
\hline
36&$ \bar{u}_{L}^{\bar{c1}} $&$\stackrel{03}{(+i)}\,\stackrel{12}{[-]}|
\stackrel{56}{[-]}\,\stackrel{78}{[-]}
||\stackrel{9 \;10}{[-]}\;\;\stackrel{11\;12}{[+]}\;\;\stackrel{13\;14}{[+]} $&-1&$-\frac{1}{2}$&0&
$-\frac{1}{2}$&$-\frac{1}{2}$&$-\frac{1}{2\,\sqrt{3}}$&$-\frac{1}{6}$&$-\frac{2}{3}$&$-\frac{2}{3}$\\
\hline
37&$\bar{d}_{R}^{\bar{c1}}$&$\stackrel{03}{(+i)}\,\stackrel{12}{(+)}|\stackrel{56}{(+)}\,\stackrel{78}{[-]}
||\stackrel{9 \;10}{[-]}\;\;\stackrel{11\;12}{[+]}\;\;\stackrel{13\;14}{[+]}$&1&$\frac{1}{2}$&
$\frac{1}{2}$&0&$-\frac{1}{2}$&$-\frac{1}{2\,\sqrt{3}}$&$-\frac{1}{6}$&$-\frac{1}{6}$&$\frac{1}{3}$\\
\hline
38&$\bar{d}_{R}^{\bar{c1}} $&$\stackrel{03}{[-i]}\,\stackrel{12}{[-]}|\stackrel{56}{(+)}\,\stackrel{78}{[-]}
||\stackrel{9 \;10}{[-]}\;\;\stackrel{11\;12}{[+]}\;\;\stackrel{13\;14}{[+]} $&1&$-\frac{1}{2}$&
$\frac{1}{2}$&0&$-\frac{1}{2}$&$-\frac{1}{2\,\sqrt{3}}$&$-\frac{1}{6}$&$-\frac{1}{6}$&$\frac{1}{3}$\\
\hline
39&$ \bar{u}_{R}^{\bar{c1}}$&$\stackrel{03}{(+i)}\,\stackrel{12}{(+)}|\stackrel{56}{[-]}\,\stackrel{78}{(+)}
||\stackrel{9 \;10}{[-]}\;\;\stackrel{11\;12}{[+]}\;\;\stackrel{13\;14}{[+]}$ &1&$\frac{1}{2}$&
$-\frac{1}{2}$&0 &$-\frac{1}{2}$&$-\frac{1}{2\,\sqrt{3}}$&$-\frac{1}{6}$&$-\frac{1}{6}$&$-\frac{2}{3}$\\
\hline
40&$\bar{u}_{R}^{\bar{c1}}$&$\stackrel{03}{[-i]}\,\stackrel{12}{[-]}|\stackrel{56}{[-]}\,\stackrel{78}{(+)}
||\stackrel{9 \;10}{[-]}\;\;\stackrel{11\;12}{[+]}\;\;\stackrel{13\;14}{[+]}$
%\stackrel{9 \;10}{[-]}\;\;\stackrel{11\;12}{[+]}\;\;\stackrel{13\;14}{[+]}$
&1&$-\frac{1}{2}$&
$-\frac{1}{2}$&0&$-\frac{1}{2}$&$-\frac{1}{2\,\sqrt{3}}$&$-\frac{1}{6}$&$-\frac{1}{6}$&$-\frac{2}{3}$\\
\hline\hline
41&$ \bar{d}_{L}^{\bar{c2}}$&$ \stackrel{03}{[-i]}\,\stackrel{12}{(+)}|
\stackrel{56}{(+)}\,\stackrel{78}{(+)}
||\stackrel{9 \;10}{(+)}\;\;\stackrel{11\;12}{(-)}\;\;\stackrel{13\;14}{[+]} $ 
&-1&$\frac{1}{2}$&0&
$\frac{1}{2}$&$\frac{1}{2}$&$-\frac{1}{2\,\sqrt{3}}$&$-\frac{1}{6}$&$\frac{1}{3}$&$\frac{1}{3}$\\
\hline
42&$\bar{d}_{L}^{\bar{c2}}$&$\stackrel{03}{(+i)}\,\stackrel{12}{[-]}|\stackrel{56}{(+)}\,\stackrel{78}{(+)}
||\stackrel{9 \;10}{(+)}\;\;\stackrel{11\;12}{(-)}\;\;\stackrel{13\;14}{[+]}$
&-1&$-\frac{1}{2}$&0&
$\frac{1}{2}$&$\frac{1}{2}$&$-\frac{1}{2\,\sqrt{3}}$&$-\frac{1}{6}$&$\frac{1}{3}$&$\frac{1}{3}$\\
\hline
43&$\bar{u}_{L}^{\bar{c2}}$&$\stackrel{03}{[-i]}\,\stackrel{12}{(+)}|\stackrel{56}{[-]}\,\stackrel{78}{[-]}
||\stackrel{9 \;10}{(+)}\;\;\stackrel{11\;12}{(-)}\;\;\stackrel{13\;14}{[+]}$
&-1&$\frac{1}{2}$&0&
$-\frac{1}{2}$&$\frac{1}{2}$&$-\frac{1}{2\,\sqrt{3}}$&$-\frac{1}{6}$&$-\frac{2}{3}$&$-\frac{2}{3}$\\
\hline
44&$ \bar{u}_{L}^{\bar{c2}} $&$\stackrel{03}{(+i)}\,\stackrel{12}{[-]}|
\stackrel{56}{[-]}\,\stackrel{78}{[-]}
||\stackrel{9 \;10}{(+)}\;\;\stackrel{11\;12}{(-)}\;\;\stackrel{13\;14}{[+]} $
&-1&$-\frac{1}{2}$&0&
$-\frac{1}{2}$&$\frac{1}{2}$&$-\frac{1}{2\,\sqrt{3}}$&$-\frac{1}{6}$&$-\frac{2}{3}$&$-\frac{2}{3}$\\
\hline
45&$\bar{d}_{R}^{\bar{c2}}$&$\stackrel{03}{(+i)}\,\stackrel{12}{(+)}|\stackrel{56}{(+)}\,\stackrel{78}{[-]}
||\stackrel{9 \;10}{(+)}\;\;\stackrel{11\;12}{(-)}\;\;\stackrel{13\;14}{[+]}$
&1&$\frac{1}{2}$&
$\frac{1}{2}$&0&$\frac{1}{2}$&$-\frac{1}{2\,\sqrt{3}}$&$-\frac{1}{6}$&$-\frac{1}{6}$&$\frac{1}{3}$\\
\hline
46&$\bar{d}_{R}^{\bar{c2}} $&$\stackrel{03}{[-i]}\,\stackrel{12}{[-]}|\stackrel{56}{(+)}\,\stackrel{78}{[-]}
||\stackrel{9 \;10}{(+)}\;\;\stackrel{11\;12}{(-)}\;\;\stackrel{13\;14}{[+]} $
&1&$-\frac{1}{2}$&
$\frac{1}{2}$&0&$\frac{1}{2}$&$-\frac{1}{2\,\sqrt{3}}$&$-\frac{1}{6}$&$-\frac{1}{6}$&$\frac{1}{3}$\\
\hline
47&$ \bar{u}_{R}^{\bar{c2}}$&$\stackrel{03}{(+i)}\,\stackrel{12}{(+)}|\stackrel{56}{[-]}\,\stackrel{78}{(+)}
||\stackrel{9 \;10}{(+)}\;\;\stackrel{11\;12}{(-)}\;\;\stackrel{13\;14}{[+]}$
 &1&$\frac{1}{2}$&
$-\frac{1}{2}$&0 &$\frac{1}{2}$&$-\frac{1}{2\,\sqrt{3}}$&$-\frac{1}{6}$&$-\frac{1}{6}$&$-\frac{2}{3}$\\
\hline
48&$\bar{u}_{R}^{\bar{c2}}$&$\stackrel{03}{[-i]}\,\stackrel{12}{[-]}|\stackrel{56}{[-]}\,\stackrel{78}{(+)}
||\stackrel{9 \;10}{(+)}\;\;\stackrel{11\;12}{(-)}\;\;\stackrel{13\;14}{[+]}$
&1&$-\frac{1}{2}$&
$-\frac{1}{2}$&0&$\frac{1}{2}$&$-\frac{1}{2\,\sqrt{3}}$&$-\frac{1}{6}$&$-\frac{1}{6}$&$-\frac{2}{3}$\\
\hline\hline
%\hline 
%$\cdots$ &&&&&&&&&&& \\
%\hline\hline
49&$ \bar{d}_{L}^{\bar{c3}}$&$ \stackrel{03}{[-i]}\,\stackrel{12}{(+)}|
\stackrel{56}{(+)}\,\stackrel{78}{(+)}
||\stackrel{9 \;10}{(+)}\;\;\stackrel{11\;12}{[+]}\;\;\stackrel{13\;14}{(-)} $ &-1&$\frac{1}{2}$&0&
$\frac{1}{2}$&$0$&$\frac{1}{\sqrt{3}}$&$-\frac{1}{6}$&$\frac{1}{3}$&$\frac{1}{3}$\\
\hline 
50&$\bar{d}_{L}^{\bar{c3}}$&$\stackrel{03}{(+i)}\,\stackrel{12}{[-]}|\stackrel{56}{(+)}\,\stackrel{78}{(+)}
||\stackrel{9 \;10}{(+)}\;\;\stackrel{11\;12}{[+]}\;\;\stackrel{13\;14}{(-)} $&-1&$-\frac{1}{2}$&0&
$\frac{1}{2}$&$0$&$\frac{1}{\sqrt{3}}$&$-\frac{1}{6}$&$\frac{1}{3}$&$\frac{1}{3}$\\
\hline
51&$\bar{u}_{L}^{\bar{c3}}$&$\stackrel{03}{[-i]}\,\stackrel{12}{(+)}|\stackrel{56}{[-]}\,\stackrel{78}{[-]}
||\stackrel{9 \;10}{(+)}\;\;\stackrel{11\;12}{[+]}\;\;\stackrel{13\;14}{(-)} $&-1&$\frac{1}{2}$&0&
$-\frac{1}{2}$&$0$&$\frac{1}{\sqrt{3}}$&$-\frac{1}{6}$&$-\frac{2}{3}$&$-\frac{2}{3}$\\
\hline
52&$ \bar{u}_{L}^{\bar{c3}} $&$\stackrel{03}{(+i)}\,\stackrel{12}{[-]}|
\stackrel{56}{[-]}\,\stackrel{78}{[-]}
||\stackrel{9 \;10}{(+)}\;\;\stackrel{11\;12}{[+]}\;\;\stackrel{13\;14}{(-)}  $&-1&$-\frac{1}{2}$&0&
$-\frac{1}{2}$&$0$&$\frac{1}{\sqrt{3}}$&$-\frac{1}{6}$&$-\frac{2}{3}$&$-\frac{2}{3}$\\
\hline
53&$\bar{d}_{R}^{\bar{c3}}$&$\stackrel{03}{(+i)}\,\stackrel{12}{(+)}|\stackrel{56}{(+)}\,\stackrel{78}{[-]}
||\stackrel{9 \;10}{(+)}\;\;\stackrel{11\;12}{[+]}\;\;\stackrel{13\;14}{(-)} $&1&$\frac{1}{2}$&
$\frac{1}{2}$&0&$0$&$\frac{1}{\sqrt{3}}$&$-\frac{1}{6}$&$-\frac{1}{6}$&$\frac{1}{3}$\\
\hline
54&$\bar{d}_{R}^{\bar{c3}} $&$\stackrel{03}{[-i]}\,\stackrel{12}{[-]}|\stackrel{56}{(+)}\,\stackrel{78}{[-]}
||\stackrel{9 \;10}{(+)}\;\;\stackrel{11\;12}{[+]}\;\;\stackrel{13\;14}{(-)} $&1&$-\frac{1}{2}$&
$\frac{1}{2}$&0&$0$&$\frac{1}{\sqrt{3}}$&$-\frac{1}{6}$&$-\frac{1}{6}$&$\frac{1}{3}$\\
\hline
55&$ \bar{u}_{R}^{\bar{c3}}$&$\stackrel{03}{(+i)}\,\stackrel{12}{(+)}|\stackrel{56}{[-]}\,\stackrel{78}{(+)}
||\stackrel{9 \;10}{(+)}\;\;\stackrel{11\;12}{[+]}\;\;\stackrel{13\;14}{(-)} $ &1&$\frac{1}{2}$&
$-\frac{1}{2}$&0 &$0$&$\frac{1}{\sqrt{3}}$&$-\frac{1}{6}$&$-\frac{1}{6}$&$-\frac{2}{3}$\\
\hline
56&$\bar{u}_{R}^{\bar{c3}}$&$\stackrel{03}{[-i]}\,\stackrel{12}{[-]}|\stackrel{56}{[-]}\,\stackrel{78}{(+)}
||\stackrel{9 \;10}{(+)}\;\;\stackrel{11\;12}{[+]}\;\;\stackrel{13\;14}{(-)} $&1&$-\frac{1}{2}$&
$-\frac{1}{2}$&0&$0$&$\frac{1}{\sqrt{3}}$&$-\frac{1}{6}$&$-\frac{1}{6}$&$-\frac{2}{3}$\\
\hline\hline
%$\cdots$ &&&&&&&&&&& \\
%\hline\hline
57&$ \bar{e}_{L}$&$ \stackrel{03}{[-i]}\,\stackrel{12}{(+)}|
\stackrel{56}{(+)}\,\stackrel{78}{(+)}
||\stackrel{9 \;10}{[-]}\;\;\stackrel{11\;12}{(-)}\;\;\stackrel{13\;14}{(-)} $ &-1&$\frac{1}{2}$&0&
$\frac{1}{2}$&$0$&$0$&$\frac{1}{2}$&$1$&$1$\\
\hline 
58&$\bar{e}_{L}$&$\stackrel{03}{(+i)}\,\stackrel{12}{[-]}|\stackrel{56}{(+)}\,\stackrel{78}{(+)}
||\stackrel{9 \;10}{[-]}\;\;\stackrel{11\;12}{(-)}\;\;\stackrel{13\;14}{(-)}$&-1&$-\frac{1}{2}$&0&
$\frac{1}{2}$ &$0$&$0$&$\frac{1}{2}$&$1$&$1$\\
\hline
59&$\bar{\nu}_{L}$&$\stackrel{03}{[-i]}\,\stackrel{12}{(+)}|\stackrel{56}{[-]}\,\stackrel{78}{[-]}
||\stackrel{9 \;10}{[-]}\;\;\stackrel{11\;12}{(-)}\;\;\stackrel{13\;14}{(-)}$&-1&$\frac{1}{2}$&0&
$-\frac{1}{2}$&$0$&$0$&$\frac{1}{2}$&$0$&$0$\\
\hline 
60&$ \bar{\nu}_{L} $&$\stackrel{03}{(+i)}\,\stackrel{12}{[-]}|
\stackrel{56}{[-]}\,\stackrel{78}{[-]}
||\stackrel{9 \;10}{[-]}\;\;\stackrel{11\;12}{(-)}\;\;\stackrel{13\;14}{(-)} $&-1&$-\frac{1}{2}$&0&
$-\frac{1}{2}$&$0$&$0$&$\frac{1}{2}$&$0$&$0$\\
\hline
61&$\bar{\nu}_{R}$&$\stackrel{03}{(+i)}\,\stackrel{12}{(+)}|\stackrel{56}{[-]}\,\stackrel{78}{(+)}
||\stackrel{9 \;10}{[-]}\;\;\stackrel{11\;12}{(-)}\;\;\stackrel{13\;14}{(-)}$&1&$\frac{1}{2}$&
$-\frac{1}{2}$&0&$0$&$0$&$\frac{1}{2}$&$\frac{1}{2}$&$0$\\
\hline
62&$\bar{\nu}_{R} $&$\stackrel{03}{[-i]}\,\stackrel{12}{[-]}|\stackrel{56}{[-]}\,\stackrel{78}{(+)}
||\stackrel{9 \;10}{[-]}\;\;\stackrel{11\;12}{(-)}\;\;\stackrel{13\;14}{(-)} $&1&$-\frac{1}{2}$&
$-\frac{1}{2}$&0&$0$&$0$&$\frac{1}{2}$&$\frac{1}{2}$&$0$\\
\hline
63&$ \bar{e}_{R}$&$\stackrel{03}{(+i)}\,\stackrel{12}{(+)}|\stackrel{56}{(+)}\,\stackrel{78}{[-]}
||\stackrel{9 \;10}{[-]}\;\;\stackrel{11\;12}{(-)}\;\;\stackrel{13\;14}{(-)}$ &1&$\frac{1}{2}$&
$\frac{1}{2}$&0 &$0$&$0$&$\frac{1}{2}$&$\frac{1}{2}$&$1$\\
\hline
64&$\bar{e}_{R}$&$\stackrel{03}{[-i]}\,\stackrel{12}{[-]}|\stackrel{56}{(+)}\,\stackrel{78}{[-]}
||\stackrel{9 \;10}{[-]}\;\;\stackrel{11\;12}{(-)}\;\;\stackrel{13\;14}{(-)}$&1&$-\frac{1}{2}$&
$\frac{1}{2}$&0&$0$&$0$&$\frac{1}{2}$&$\frac{1}{2}$&$1$\\
\hline 
\end{supertabular}
}
\end{center}
%\end{small}

%\end{tiny}
%

In Ref.~\cite{EPJC2017} (and the references therein) it is demonstrated that all the vector 
gauge fields, which are the gauge fields of the charges presented in 
Table~\ref{Table SMandSCFTspinors.}, 
origin in spin connections, that is in gravity~\footnote{To unify and quantize the vector (as well as 
the scalar) gauge fields, which (both, scalar and vector ones) originate in gravity, the quantization of 
gravity should be known. We assume in this paper that all the vector gauge fields are quantized as 
$U(1), SU(2)$ ($SU(2)_{I}, SU(2)_{II}$ indeed) and $SU(3)$ vector gauge fields.}. 
The gravitational origin of these vector gauge fields can be neglected as long as they are weak in 
comparison with fields which make ($d-4$) space curled. In this sense the quantization of the vector 
gauge fields of the groups $U(1), SU(2)_{I}$, $SU(2)_{II}$ and $SU(3)$, although originating in  
$SO(13,1)$, can from the point of view of the flat $d=(3+1)$ space still be helpful  for understanding 
the {\it standard model} "miraculous" triangle anomaly cancellation.

% Mention  Norma /Holger Bled proceeding contribution of the cut off possibility.

Let be mentioned that  the {\it spin-charge-family} theory  is able --- while starting from a very 
simple action in $d = (13+1)$,  Eq.~(\ref{wholeaction}), with massless fermions with the spin of 
the two kinds, Eqs.~(\ref{gammatildegamma},\ref{sabtildesab}), (one kind taking care of the spin 
and the charges of the family members in $d=(3+1)$, the second kind taking care of the families 
and family charges, the two kinds corresponding to the left and to the right 
multiplication~\cite{JMP2015} of any Clifford algebra object,  respectively) coupled only to the 
gravity (through the vielbeins and the two kinds of the corresponding spin connections) --- to 
explain not only all the assumptions of the {\it standard model}, 
but answers also several open questions beyond the 
{\it standard model}~\cite{norma2014MatterAntimatter,JMP2015,norma2016higgs}.  

In Sect.~\ref{anomalycancellation} the contribution to the triangle anomalies of the {\it standard
model} massless fermions and anti-fermions are discussed from the point of view that the $SO(3,1), 
SU(2), SU(3)$ and $U(1)$ are the subgroups of the orthogonal group $SO(13,1)$.

We comment in Subsect.~\ref{so10} the advantage of embedding the {\it standard mode}l groups 
into the orthogonal group $SO(13,1)$ rather than into $SO(10)$ group.

We briefly present in Sect.~\ref{SCFT} of Appendix those main points of the {\it spin-charge-family} 
theory, which are needed in this paper. For more informations we advise the reader to read 
Refs.~\cite{JMP2015,norma2014MatterAntimatter,NBled2013,NBled2012,IARD2016} and the
 references therein.

In Subsect.~\ref{explicit} we present in details in two examples 
how can one extract from the products of nilpotents and projectors, representing a state in the 
fundamental representation of $SO(13,1)$, properties of the spinor described by this state. This is
done in {\it Example 1.} and {\it Example 2}.

 We also demonstrate in {\it Example 3.}  how the family states can be reached from any of the 
members presented in Table~\ref{Table so13+1.}. 

In {\it Example 4.} the $SO(10)$ content is described in details from the point of view of the 
$SO(13,1)$ fundamental representation.

In App.~\ref{technique} the technique to represent spinors, used in this paper, is briefly explained.

\section{Standard model triangle anomaly cancellation looks natural if $SO(3,1), SU(2), SU(3)$ and 
$U(1)$ are embedded in $SO(13,1)$}
\label{anomalycancellation}

The cancellation of the triangle anomalies of the massless quarks and leptons and their anti-particles 
looks from the point of view of the {\it standard model} "miraculous".

The triangle anomaly  of the {\it standard model}  occurs if the traces in Eq.(\ref{anomaly0}) are not
zero for either the left handed quarks and leptons and anti-quarks and anti-leptons or the right handed 
quarks and leptons and anti-quarks and anti-leptons for the Feynman triangle diagrams in which the 
gauge vector fields of the type 
\begin{eqnarray}
\label{anomalyparticular}
&& U(1)\times U(1) \times U(1)\, ,\nonumber\\ 
&& SU(2)\times SU(2) \times U(1)\, ,\nonumber\\
&& SU(3)\times SU(3) \times SU(3)\, ,\nonumber\\ 
&& SU(3)\times SU(3) \times U(1)\, ,\nonumber\\  
&& U(1) \times {\rm gravitational}\,  
\end{eqnarray}
contribute to the triangle anomaly.

Let us first in Table~\ref{Table SMandSCFTspinors.} present by the {\it standard model} assumed 
properties of the members of any one massless family, running in the triangle. The corresponding 
data are represented in the first eight columns (up to $||$). The last two columns, taken from 
Table~\ref{Table so13+1.}, describe additional properties which quarks and leptons and anti-quarks 
and anti-leptons would have, if the {\it standard model} groups $SO(3,1)$, $SU(2), SU(3)$ and 
$U(1)$ are embedded into the $SO(13,1)$ group.  We shall comment these last two columns later.
In the {\it spin-charge-family} theory the family quantum numbers are determined by the second
kind of the Clifford algebra objects $\tilde{S}^{ab}$, which commute with $S^{ab}$ describing
spins and charges. Correspondingly the spins and charges 
are the same for all the families.
\begin{table}
{\tiny%
\begin{center}
\begin{tabular}{|r r | c r r r r r ||r r|}
\hline
           & & hand-   & weak    & hyper  & colour&charge & elm   & $SU(2)_{II}$&$U(1)_{II}$\\
            &&edness  & charge  & charge &        &          &charge&     charge     &  charge     \\
$i_{L}$&name    &$ \Gamma^{(3,1)}$&$ \tau^{13}$  & $ Y$ & $\tau^{33}$&$\tau^{38}$
&$Q$&$\tau^{23}$&$\tau^{4}$\\
\hline
$1_{L}$&$ u_{L} $&$ -1     $&$\frac{1}{2}$ &$ \frac{1}{6}$&$\frac{1}{2} $&$\frac{1}{2\sqrt{3}}$&
$ \frac{2}{3}$&0&$\frac{1}{6}$\\
$2_{L}$&$ d_{L} $&$ -1    $&$ -\frac{1}{2}$ &$ \frac{1}{6}$&$\frac{1}{2}$&$\frac{1}{2\sqrt{3}}$ &
$-\frac{1}{3}$&0&$\frac{1}{6}$\\
$3_{L}$&$ u_{L} $&$ -1     $&$\frac{1}{2}$ &$ \frac{1}{6}$&$-\frac{1}{2} $&$\frac{1}{2\sqrt{3}}$&
$ \frac{2}{3}$&0&$\frac{1}{6}$\\
$4_{L}$&$ d_{L} $&$ -1    $&$ -\frac{1}{2}$ &$ \frac{1}{6}$&$-\frac{1}{2}$&$\frac{1}{2\sqrt{3}}$ &
$-\frac{1}{3}$&0&$\frac{1}{6}$\\
$5_{L}$&$ u_{L} $&$ -1     $&$\frac{1}{2}$ &$ \frac{1}{6}$&$ 0 $&$-\frac{1}{\sqrt{3}}$&
$ \frac{2}{3}$&0&$\frac{1}{6}$\\
$6_{L}$&$ d_{L} $&$ -1    $&$ -\frac{1}{2}$ &$ \frac{1}{6}$&$ 0$&$-\frac{1}{\sqrt{3}}$ &
$-\frac{1}{3}$&0&$\frac{1}{6}$\\
\hline
$7_{L}$&$\nu_{L} $&$ -1  $&$ \frac{1}{2}$&$ -\frac{1}{2}$& 0&0&  0  &$0$&$ -\frac{1}{2}$  \\
$8_{L}$&$ e^{L}  $&$ -1  $&$-\frac{1}{2}$&$ -\frac{1}{2}$& 0&0&$-1$&$0$&$ -\frac{1}{2}$ \\
\hline \hline
%anti-particles
$9_{L}$&$ \bar{u}{L} $&$ -1$&$  0 $ &$ -\frac{2}{3}$& $ - \frac{1}{2}$ &$-\frac{1}{2\sqrt{3}}$
&$- \frac{2}{3}$ &$ - \frac{1}{2}$ &$ - \frac{1}{6}$\\
$10_{L}$&$ \bar{d}{L} $&$ -1$&$  0 $ &$  \frac{1}{3}$& $ - \frac{1}{2}$ &$-\frac{1}{2\sqrt{3}}$
&$  \frac{1}{3}$ &$   \frac{1}{2}$ &$ - \frac{1}{6}$\\
$11_{L}$&$ \bar{u}{L} $&$ -1$&$  0 $ &$ -\frac{2}{3}$& $  \frac{1}{2}$ &$-\frac{1}{2\sqrt{3}}$
&$- \frac{2}{3}$ &$ - \frac{1}{2}$ &$ - \frac{1}{6}$\\
$12_{L}$&$ \bar{d}{L} $&$ -1$&$  0 $ &$  \frac{1}{3}$& $  \frac{1}{2}$ &$-\frac{1}{2\sqrt{3}}$
&$  \frac{1}{3}$ &$   \frac{1}{2}$ &$ - \frac{1}{6}$\\
$13_{L}$&$ \bar{u}{L} $&$ -1$&$  0 $ &$ -\frac{2}{3}$& $ 0 $&$ \frac{1}{\sqrt{3}}$
&$- \frac{2}{3}$ &$ - \frac{1}{2}$ &$ - \frac{1}{6}$\\
$14_{L}$&$ \bar{d}{L} $&$ -1$&$  0 $ &$  \frac{1}{3}$& $ 0 $&$\frac{1}{\sqrt{3}}$
&$  \frac{1}{3}$ &$   \frac{1}{2}$ &$ - \frac{1}{6}$\\
\hline
$15_{L}$&$\bar{\nu}_{L}$&$ -1$&$ 0 $&$ 0 $& 0&0&$0$&$- \frac{1}{2}$&$ \frac{1}{2}$  \\
$16_{L}$&$\bar{e}_{L}   $&$ -1$&$ 0 $&$ 1 $& 0&0&$1$&$  \frac{1}{2}$&$ \frac{1}{2}$  \\
\hline \hline\hline
%particles
$1_{R}$&$u_{R}   $&$ 1$& $0$ & $  \frac{2}{3}$&$ \frac{1}{2}$ &$\frac{1}{2\sqrt{3}}$
&$ \frac{ 2}{3}$&$ \frac{1}{2}$&$ \frac{1}{6}$ \\
$2_{R}$&$d_{R}   $&$ 1$& $0$ & $ -\frac{1}{3}$&$ \frac{1}{2}$ &$\frac{1}{2\sqrt{3}}$ 
&$-\frac{1}{3}$&$- \frac{1}{2}$&$ \frac{1}{6}$  \\
$3_{R}$&$u_{R}   $&$ 1$& $0$ & $  \frac{2}{3}$&$- \frac{1}{2}$ &$\frac{1}{2\sqrt{3}}$
&$ \frac{ 2}{3}$&$ \frac{1}{2}$&$ \frac{1}{6}$ \\
$4_{R}$&$d_{R}   $&$ 1$& $0$ & $ -\frac{1}{3}$&$- \frac{1}{2}$ &$\frac{1}{2\sqrt{3}}$ 
&$-\frac{1}{3}$&$- \frac{1}{2}$&$ \frac{1}{6}$  \\
$5_{R}$&$u_{R}   $&$ 1$& $0$ & $  \frac{2}{3}$&$ 0  $ &$-\frac{1}{\sqrt{3}}$
&$ \frac{ 2}{3}$&$ \frac{1}{2}$&$ \frac{1}{6}$ \\
$6_{R}$&$d_{R}   $&$ 1$& $0$ & $ -\frac{1}{3}$&$ 0  $ &$-\frac{1}{\sqrt{3}}$ 
&$-\frac{1}{3}$&$- \frac{1}{2}$&$ \frac{1}{6}$  \\
\hline
$7_{R}$&$\nu_{R}$&$ 1$& $0$  & $ 0$&0&0&   0 &$  \frac{1}{2}$&$- \frac{1}{2}$           \\
$8_{R}$&$ e_{R}  $&$ 1$& $0$  & $-1$&0&0&$-1$&$- \frac{1}{2}$&$- \frac{1}{2}$  \\
\hline\hline
%anti-particles%%%%12.08.2016 8:43%%%%12.08.2016 8:43
$9_{R}$&$\bar{u}_{R}   $&$ 1$& $ -\frac{1}{2}$ & $ - \frac{1}{6}$&$ -\frac{1}{2}$ &$-\frac{1}{2\sqrt{3}}$
&$ - \frac{ 2}{3}$&0& $ - \frac{1}{6}$\\
$10_{R}$&$\bar{d}_{R}   $&$ 1$&$  \frac{1}{2}$ & $  - \frac{1}{6}$&$ -\frac{1}{2}$ &$-\frac{1}{2\sqrt{3}}$
&$    \frac{1}{3}$&0& $ - \frac{1}{6}$\\
$11_{R}$&$\bar{u}_{R}   $&$ 1$& $ -\frac{1}{2}$ & $ -  \frac{1}{6}$&$ \frac{1}{2}$ &$-\frac{1}{2\sqrt{3}}$
&$ - \frac{ 2}{3}$&0& $ - \frac{1}{6}$\\
$12_{R}$&$\bar{d}_{R}   $&$ 1$&$  \frac{1}{2}$ & $ -  \frac{1}{6}$&$  \frac{1}{2}$ &$-\frac{1}{2\sqrt{3}}$
&$    \frac{1}{3}$&0& $ - \frac{1}{6}$\\
$13_{R}$&$\bar{u}_{R}   $&$ 1$& $ -\frac{1}{2}$ & $ -  \frac{1}{6}$&$ 0 $ &$ \frac{1}{\sqrt{3}}$
&$ - \frac{ 2}{3}$&0& $ - \frac{1}{6}$\\
$14_{R}$&$\bar{d}_{R}   $&$ 1$&$  \frac{1}{2}$ & $  -  \frac{1}{6}$&$ 0 $ &$ \frac{1}{\sqrt{3}}$
&$    \frac{1}{3}$&0& $ - \frac{1}{6}$\\
\hline
$15_{R}$&$\bar{\nu}_{R}$&$ 1$& $ -\frac{1}{2}$ & $  \frac{1}{2}$&0&0&$0$&0&$\frac{1}{2}$        \\
$16_{R}$&$\bar{e}_{R}   $&$ 1$& $  \frac{1}{2}$ & $  \frac{1}{2}$&0&0&$1$&0&$\frac{1}{2}$ \\
\hline\hline
\end{tabular}
  \end{center}
%  
%\caption{\label{Table I.}
}
\caption{\label{Table SMandSCFTspinors.} 
\tiny{Properties  of the left handed quarks and leptons and 
of the left handed anti-quarks and anti-leptons (the first $16$ lines)  and of the right handed quarks 
and leptons and the right handed anti-quarks and anti-leptons (the second $16$ lines), as assumed by 
the {\it standard model}, are presented in the first seven columns. In the last two columns the two 
quantum numbers are added, which fermions and anti-fermions would have if the {\it standard model} 
groups $SO(3,1)$, $SU(2)$, $SU(3)$ and $U(1)$ are embedded into the $SO(13,1)$ group. 
All the infinitesimal generators of the subgroups of the orthogonal group $SO(13,1)$, representing 
charges in $d=(3+1)$, are in the {\it spin-charge-family} theory the superposition of the generators 
$S^{st}$, Eqs.~(\ref{so42},\ref{so64}). The handedness is defined in Eq.(\ref{hand}). It is 
demonstrated in {\it Example 1.} and {\it Example 2.} of Subsect.~\ref{explicit} how can one read 
handedness 
and charges from products of nilpotents and projectors, which in Table~\ref{Table so13+1.} determine 
spinor states, representing quarks and leptons and anti-quarks and anti-leptons.
% The weak charge operator is
%$\vec{\tau}^1= \frac{1}{2} (S^{58}- S^{67}, $ $S^{57}+ S^{68}, S^{56}- S^{78})$, the second
%$SU(2)_{II}$ operator is $\vec{\tau}^2= $ $\frac{1}{2}$ $ (S^{58}+ S^{67}$, $S^{57}- S^{68}, 
%S^{56}+ S^{78})$, the colour charge operator is $\vec{\tau}^{3}=$ $ \frac{1}{2}(S^{9\,12}- $ $
%S^{10\,11}$, $S^{9\,11}+ S^{10\,12}$, $S^{9\,10}- S^{11\,12}$, $S^{9\,14}- S^{10\,13}$,
%$S^{9\,13}+ S^{10\,14}$, $S^{11\,14}- S^{12\,13}$, $S^{11\,13}+ 
%S^{12\,14}$, $\frac{1}{\sqrt{3}} (S^{9\,10}+ S^{11\,12} - $ $2S^{13\,14})$) and the $U(1)_{II}$ 
%operator is equal to $\tau^{4}=$ $-\frac{1}{3} $ $(S^{9\,10}+ S^{11\,12}$ $+ S^{13\,14})$ and 
%represents the "spinor" charge. Correspondingly  the hyper charge $Y$ is the sum of ($\tau^{4} + $ $
%\tau^{23}$). 
%One can read all these properties from Table~\ref{Table so13+1.}, or even calculate from this table, 
%if taking into account Eq.~(\ref{grapheigen}).  Some examples are presented in Subsect.~\ref{explicit}. 
The whole quark part appears, due to the colour charges, three times.  
These quantum numbers are the same for all the families. % also in the {\it spin-charge-family} theory.
}
}
% \end{table}
\end{table}

To calculate the traces required in Eq.~(\ref{anomaly0}) for the triangle anomalies of 
Eq.~(\ref{anomalyparticular}) the quantum numbers of the left handed spinors and anti-spinors, as 
well as of the right handed spinors and anti-spinors, presented in Table~\ref{Table SMandSCFTspinors.},
are needed.

Before demonstrating the well known results that the required traces are in the {\it standard model} 
equal to zero, assuring the "miraculous" anomaly free triangle diagrams, let us comment the quantum 
numbers of the same spinors, if the {\it standard model} subgroups --- $SO(3,1)$, $SU(2)$, $SU(3)$, 
$U(1)$ --- are embedded into $SO(13,1)$.
In this case each of the {\it standard model} family member (of any family) has additional quantum 
numbers to the quantum numbers assumed by  the {\it standard model}, as one can see in 
Table~\ref{Table so13+1.}, where the quarks and leptons, left and right handed, and the corresponding 
anti-quarks and anti-leptons, are presented as product of nilpotents and projectors, out of which one can 
easily read the quantum numbers of states, as presented in {\it Example 1.} and {\it Example 2.} in 
Subsect.~\ref{explicit}.

The desired (from the point of view of the observed properties of quarks and leptons) subgroups of 
the $SO(13,1)$ group are $SO(7,1) \times SO(6)$. The desired subgroups of $SO(6)$ are the 
colour group $SU(3)$ with the generators denoted by $\tau^{3i}, \, i=1,\dots,8$, Eq.~(\ref{so64}), 
and the group $U(1)$ (we shall call it $U(1)_{II}$)  with the generator $\tau^{4}$, Eq.~(\ref{so64}). 

The desired subgroups of $SO(7,1)$ are $SO(3,1) \times SU(2)_{I}\times SU(2)_{II}$. This kind 
of breaking symmetries makes handedness of quarks and leptons connected with the weak charge
($\vec{\tau}^{1}$, Eq.~(\ref{so42}),  $ SU(2)_{I}$),  and the hyper charge $Y$  
(Eqs.~(\ref{Y},\ref{so42},\ref{so64}), originated in $ SU(2)_{II}$ and $U(1)_{II}$).

All these generators are expressed 
in term of the generators $S^{ab}$ ($\tau^{Ai} = \sum_{a,b} c^{Ai}{}_{ab} \, S^{ab}$) of the 
orthogonal group $SO(13,1)$.
One sees in Table~\ref{Table so13+1.}  and in Table~\ref{Table SMandSCFTspinors.} that all the 
quarks have (the "fermion charge") $\tau^{4}=\frac{1}{6}$, the anti-quarks have 
$\tau^{4}=-\frac{1}{6}$, while the leptons  have $\tau^{4}=- \frac{1}{2}$ and the 
anti-leptons have $\tau^{4}= \frac{1}{2}$. Correspondingly the trace of $\tau^{4}$ over all the 
family members of one handedness is equal to zero. Let us remember that in one  Weyl (spinor) 
representation of the $SO(13,1)$ group the right handed neutrino (and correspondingly  
anti-neutrino) is the regular member of the representation. Since all the spinor states are represented by 
nilpotents and projectors which are the eigenstates of $S^{03}, S^{1 2}$, $ S^{56}$, $ S^{78}$,
$ S^{9\,10}$, $ S^{11\,12}$ and $ S^{13\,14}$, it follows that also $ \tau^{13}$, 
$\tau^{23}$, $\tau^{33}$, $\tau^{38}$ and $\tau^{4}$ are diagonal in this basis. Correspondingly
it is easy to check all the quantum numbers presented in Table~\ref{Table so13+1.} and in 
Table~\ref{Table SMandSCFTspinors.}  if taking into account Eq.~(\ref{grapheigen}) and the 
expressions for  $\tau^{Ai} = \sum_{a,b} c^{Ai}{}_{ab} \, S^{ab}$, presented in Eqs.~(\ref{so42},%
\ref{so64},\ref{Y}).

The representations of the subgroup of $SO(7,1)$ (the part of nilpotents and projectors up to the 
sign $||$ of the eight vectors) are the same for all the quarks and the leptons, and the same for all 
the anti-quarks and anti-leptons, as seen in Table~\ref{Table so13+1.}. $SO(7,1)$ has as subgroups 
$SO(3,1)$, $SU(2)_{I}$, $SU(2)_{II}$, with the generators $\vec{N}_{\pm}$ (Eq.~(\ref{so1+3})),
$\vec{\tau}^{1}$ and $\vec{\tau}^{2}$  (Eq.~(\ref{so42})). The group $SO(3,1)$ determines the 
spin and handedness of spinors (and anti-spinors).
The left handed spinors are $ SU(2)_{I}$ (weak) doublets and $SU(2)_{II}$ singlets, while the right 
handed spinors  are the $ SU(2)_{I}$ (weak) singlets and the $SU(2)_{II}$ doublets. Correspondingly
the left handed anti-spinors are  the $ SU(2)_{I}$ (weak) singlets and the $SU(2)_{II}$ doublets, 
while the right  handed anti-spinors  are the $ SU(2)_{I}$ (weak) doublets and the $SU(2)_{II}$ 
singlets.

Quarks distinguish from leptons (as well as anti-quarks from anti-leptons) only in the 
$SU(3) \times U(1)_{II}$ part of the state. 

The hyper charge of the {\it standard model}  corresponds to the sum of  $\tau^{4}$ and 
$\tau^{23}$
\begin{eqnarray}
\label{Y}
Y&=& \tau^{4} + \tau^{23} \,,
\end{eqnarray}
and manifests as the conserved charge after the $SU(2)_{II}$ symmetry is 
broken~\cite{JMP2015,norma2014MatterAntimatter,norma92,norma93,norma94,pikanorma,gmdn07,%
gn,gn2013,gn2015,NPLB,N2014scalarprop,IARD2016,EPJC2017}, leaving the hyper charge $Y$ unbroken. 

A short presentation of the properties of all the members of one family, appearing in the Weyl 
(fundamental) representation of $SO(13,1)$ (Table~\ref{Table so13+1.}) 
is made in Table~\ref{Table SMandSCFTspinors.}, if also the last two columns are taken into account. 
Looking at this table one easily recognizes where does the "miraculous" cancellation of the triangle 
anomalies emerge. That a "miraculous" cancellation of triangle anomalies appear to be "trivial", if one 
takes into account that the {\it standard model} groups have the origin in $SO(13,1)$, supports the 
{\it spin-charge-family} theory as a right next step beyond the {\it standard model}.

It is worthwhile to point out again that within the one Weyl representation of $SO(13,1)$, when 
$SO(13,1)$ breaks first into $SO(7,1) \times SU(3) \times U(1)_{II}$ (and then further into
$SO(3,1) \times $ $SU(2) \times U(1) \times SU(3)$),
Table~\ref{Table so13+1.}, the handedness of particles as well as of anti-particles is uniquely 
connected with their charges, while in $SO(10)$, for example, such a connection must be "done by
hand". 

The traces of Eq.~(\ref{anomalyparticular}) for possible anomalous triangle diagrams are calculated
in Subsect.~\ref{traces}.

\subsection{Traces of the left handed spinors and the left handed anti-spinors and of the right handed 
spinors and the right handed anti-spinors of one family of quarks and leptons}
\label{traces}

Let us calculate the traces, Eq.~(\ref{anomalyparticular}), for possible anomalous triangle diagrams in 
$d=(3+1)$. 
One must evaluate the trace of the product  of three generators and sum the trace over all the states
of either the left handed members ---  $16$ states presented in the first part of 
Table~\ref{Table SMandSCFTspinors.} --- or the right handed members --- $16$ states presented in 
the second part of Table~\ref{Table SMandSCFTspinors.}.
Let us recognize again that in the case of embedding the {\it standard model} groups into $SO(13,1)$
we have $Y=(\tau^{4} + \tau^{23}) $. 

For the triangle Feynman diagram,  to which three hyper $U(1)$ boson fields contribute, we must 
evaluate  $\sum_{i} Tr (Y_{i})^3$, in which the sum runs over all the members ($i=(1,..,16)$) of the 
left handed spinors and anti-spinors, and of the right handed spinors and anti-spinors separately. 
In the case of embedding the {\it standard model} groups into $SO(13,1)$ we have
\begin{eqnarray}
\label{3Ytriangle}
\sum_{i_{L,R}}\, (Y_{i_{L,R}})^3&=&\sum_{i_{L,R}}\,
 (\tau^{4}_{i_{L,R}} + \tau^{23}_{i_{L,R}})^{3}  
\nonumber\\
 &=&\sum_{i_{L,R}}\, (\tau^{4}_{i_{L,R}})^{3} + \sum_{i_{L,R}}\, (\tau^{23}_{i_{L,R}})^{3}
\nonumber\\
&+&\sum_{i_{L,R}}\, 3 \cdot (\tau^{4}_{i_{L,R}})^{2} \cdot \tau^{23}_{i_{L,R}}
  +  \sum_{i_{L,R}}\, 3 \cdot \tau^{4}_{i_{L,R}} \cdot  (\tau^{23}_{i_{L,R}})^{2}\,,
\end{eqnarray}
for either the left, $i_{L}$, or the right, $i_{R}$, handed members. 
Table~\ref{Table SMandSCFTspinors.}  demonstrates  clearly that $ (Y_{i_{L,R}})^3=0$ without 
really making any algebraic evaluation. Namely, the last column of Table~\ref{Table SMandSCFTspinors.}  
manifests that $ \sum_{i_{L}}\,
 (\tau^{4}_{i_{L}})^{3} =0 $ [in details: $\sum_{i_{L}}\, (\tau^{4}_{i_{L}})^{3}  $ 
$ = 2 \cdot 3 \cdot (\frac{1}{6})^{3} + 2 \cdot 3 \cdot (-\frac{1}{6})^{3} +2 \cdot (-\frac{1}{2})^{3} +
 2 \cdot (\frac{1}{2})^{3}=0  $]. %$=\sum_{i_{R}}\,  (\tau^{4}_{i_{R}})^{3} =0$]. 
Table~\ref{Table SMandSCFTspinors.}  also demonstrates  (the last but one column) that 
$ \sum_{i_{L}}\, (\tau^{23}_{i_{L}})^{3} =0 $ $[=(3+1) \cdot ((-\frac{1}{2})^{3}  +
 (\frac{1}{2})^{3}))]$, and that also
 $ \sum_{i_{R}}\,  (\tau^{23}_{i_{R}})^{3} =0 $ $[= (3+1) \cdot ((\frac{1}{2})^{3} +
 (- \frac{1}{2})^{3})]$. 

From Table~\ref{Table SMandSCFTspinors.}  one sees also (without calculating) that 
$\sum_{i_{L}}\, 3 \cdot (\tau^{4}_{i_{L}})^{2} \cdot \tau^{23}_{i_{L}} =0$, in particular 
$\sum_{i_{L}}\, 3 \cdot (\tau^{4}_{i_{L}})^{2} \cdot \tau^{23}_{i_{L}} $ 
$=3.\{ ((\frac{1}{2})^2 \cdot (-\frac{1}{2} +\frac{1}{2})+ 3 \cdot (-\frac{1}{6})^2  
\cdot (-\frac{1}{2} +\frac{1}{2})\}$, as well as  that $\sum_{i_{R}}\, 3 \cdot 
(\tau^{4}_{i_{R}})^{2} \cdot \tau^{23}_{i_{R}} =0$ [$=3\cdot\{ ((-\frac{1}{2})^2 
\cdot (\frac{1}{2} +
(-\frac{1}{2}))+ 3 \cdot (\frac{1}{6})^2  \cdot (\frac{1}{2} + (-\frac{1}{2}))\}$].

That the last term in Eq.~(\ref{3Ytriangle}) is zero for either the left or the right handed spinors can
also easily be seen just by looking at Table~\ref{Table SMandSCFTspinors.} [or in details: 
$ \sum_{i_{L}}\, 3 \cdot \tau^{4}_{i_{L}} \cdot  (\tau^{23}_{i_{L}})^{2}=0$
$=3\cdot \{(\frac{1}{2} ((\frac{1}{2})^2 +(-\frac{1}{2})^2) + 3\cdot (-\frac{1}{6})
 ((\frac{1}{2})^2 +(-\frac{1}{2})^2))\} $, as well as that 
$ \sum_{i_{R}}\, 3 \cdot \tau^{4}_{i_{L}} \cdot  (\tau^{23}_{i_{R}})^{2}=0$
$=3\cdot \{(-\frac{1}{2} ((\frac{1}{2})^2 +(-\frac{1}{2})^2) + 3\cdot (\frac{1}{6})
 ((\frac{1}{2})^2 +(-\frac{1}{2})^2))\} $].

Since all the members belong to one spinor representation, it is straightforwardly that all the triangle 
traces are zero, if the {\it standard model}  groups are the subgroups of the orthogonal group 
$SO(13,1)$. 

There is no need for a detailed calculations, since a look in 
Table~\ref{Table SMandSCFTspinors.} gives immediately the answer.

From only the {\it standard model} assumptions point of view the cancellation of the triangle anomalies
does look  miraculously. For our $\sum_{i_{L,R}}\, (Y_{i_{L,R}})^3$ one obtains for the left handed
members: [$3 \cdot2 \cdot (\frac{1}{6})^3 + 2 \cdot (-\frac{1}{2})^3 + 3 \cdot  ((-\frac{2}{3})^3 
+ (\frac{1}{3})^3)
+ 1^3)$ ], and for the right handed members:  [$3 \cdot ((\frac{2}{3})^3 + (-\frac{1}{3})^3)+
(- 1)^3) +3 \cdot 2 \cdot (-\frac{1}{6})^3 + 2 \cdot (\frac{1}{2})^3 $ ].

\subsection{The $SO(13,1)$ breaking into $SO(7,1) \times SO(6)$ connects handedness with the 
weak and the hyper charge while the $SO(13,1)$ breaking into $SO(10) \times SO(3,1)$ does not }
\label{so10}

Looking at Table~\ref{Table so13+1.} one immediately sees that also $SO(10)$ offers all the
degrees of freedom needed to explain the triangle anomaly free {\it standard 
model} properties, provided that connection between handedness and charges is made {\it by
assumption}. The $SO(10)$ fundamental representation of states can in our technique be 
obtained by omitting the terms describing spin and handedness of quarks and leptons, and their 
anti-particles: The first two nilpotents or projectors up to the vertical line $"|"$ must be omitted. 
Embedding this group into $SO(13,1)$ adds $SO(3,1)$ to $SO(10)$, but without the connections 
between spin and handedness and the charges.  

In Subsect.~\ref{explicit}, {\it Example 4.}, the detailed embedding of $SO(10)$ into $SO(13,1)$ is 
discussed. The choice of connecting handedness with the weak and hyper charges is there assumed.
The $SO(10)$ {\it representation alone neither connects the weak and the hyper charge with 
the handedness nor it connects the left handed states with the right handed anti-states} 
({\it or the right 
handed states with the left handed anti-states}). The weak and hyper charges can in $SO(10)$ be 
connected with the handedness only "by hand", like  the {\it standard model} does.

The $SO(13,1)$ break into $SO(7,1) \times SU(3) \times U(1)$ (and then breaking $SO(7,1)$ further
into $SO(3,1) \times SU(2) \times SU(2)$) explains all the assumptions of the standard model:\\% use footnote
{\bf i.} $\;\;$ The {\it weak charge} (appearing from $SU(2)_{I}$, which
is the subgroup of $SO(4)$) and the {\it hyper charge} (appearing as a sum of the third component 
of $SU(2)_{II}$, which is also the subgroup of  the same $SO(4)$, and of the "fermion charge" 
$U(1)_{II}$, appearing from $SO(6)$  together with $SU(3)$, after the break of the $SU(2)_{II}$ 
group, which leaves the hyper charge $Y$
 ($Y= (\tau^{23}+ \tau^4)$), Eqs.~(\ref{so42},\ref{so64}),  
%$($= \frac{1}{2} (S^{56} + S^{78}) -\frac{1}{3} (S^{\; 10}+ 
%S^{11\; 12} +  S^{13\; 14}))$ 
conserved) {\it are uniquely connected with the handedness}.\\
{\bf ii.} $\;\;$ The representation demonstrates that the {\it spinor states are up to the
$SU(3) \times U(1)_{II}$ part} (that is up to the vertical  line $"||"$ in 
Table~\ref{Table so13+1.}) the {\it same for either quarks or leptons}. 
The same is true also for anti-quarks and anti-leptons.\\
{\bf iii.} $\;\;$ Also the {\it anti-particle states are the members of the same representation}, 
what makes the triangle anomaly cancellation in $d=(3+1)$ transparent.

Correspondingly, embedding the {\it standard model} charge groups together with the spin into
$SO(13,1)$ makes not only the weak and the hyper charges of spinors connected with the
handedness and the relation among the particle states and anti-particle states transparent, it also 
explains all the assumptions of the {\it standard model} concerning charges of one family of 
quarks and leptons. This does even more~\cite{EPJC2017, JMP2015,norma2014MatterAntimatter,%
IARD2016}
%norma92,norma93,norma94,pikanorma,gmdn07,gn,gn2013,gn2015,NPLB,N2014scalarprop,
%IARD2016}
by offering the explanation for the origin of the vector gauge fields ($A^{Ai}_{m}$,
$x^m= (x^0,\cdots , x^3)$) and of the scalar gauge fields
 ($A^{Ai}_{s}$, $x^s= (x^5,\cdots , x^d)$, to which both kinds of the spin 
connection fields contribute).

%Since the {\it spin-charge-family} theory does not neglect that there are two kinds of spins, and that 
%the second kind manifest the family quantum numbers if the first kinds manifests as spin and charges, 
%the spin connections of both kinds explain the origin of the scalar fields which determine masses of 
%quark and leptons, the Yukawa coupling of the {\it standard model}, and the masses of the heavy 
%(weak) bosons.

%
\section{Conclusions}
\label{discussions}

The assumptions of the {\it standard model} (about the properties of the family members, about 
the families, about the origin of the vector gauge fields and about the origin of the Higgs's scalar and 
the Yukawa couplings) 
certainly call for the explanation, and so does the fact that the {\it standard model} 
(with all the assumed properties) works so far so (extremely) well.

The {\it spin-charge-family} theory does offer the explanation for all the assumptions
of the {\it standard model}, and beyond the {\it standard model}, making several 
predictions~\cite{EPJC2017,JMP2015,norma2014MatterAntimatter,NBled2013,NBled2012,IARD2016,%
pikanorma,portoroz03,norma92,norma93,norma94,norma95,gmdn07,gn,gn2013,gn2015,NPLB,%
N2014scalarprop}, provided that the vector and the scalar gauge fields, which origin in the spin 
connection fields --- that is in gravity (Ref.~\cite{EPJC2017}, and the references therein) --- are  
weak in comparison with fields which make ($d-4$) space curled. If the gravitational origin of these 
vector gauge fields is neglected (and correspondingly the knowledge how to quantize gravity is not 
needed), the vector gauge fields can be quantized as $U(1)$ or $SU(n)$ vector gauge fields, 
as it is done in the {\it standard model}. 
%Yet the fact that embedding the {\it standard model} groups into $SO(13,1)$  help to 
%understand the "miraculous" anomaly cancellation.
% Mention Holger Norma Bled proceeding contribution of the cut off possibility.

In this sense the {\it standard model} triangle "miraculous"  anomaly cancellation  can trivially 
be explained within the {\it spin-charge-family} theory~\cite{IARD2016},
Sect.~\ref{anomalycancellation}. 
One Weyl representation of $SO(13,1)$ (the model starts with the simple action 
(Eq.~(\ref{wholeaction})) 
in $d=(13+1)$, leading in the low energy regime to the {\it standard model} action with the right 
handed neutrino and its antineutrino included, explaining the appearance of the families, of the 
scalar field and the Yukawa couplings) namely contains, if analyzed from the point of view of the
{\it standard model} groups ($SO(3,1), SU(2)$, $SU(3)$ and $U(1)$),  all the quarks and leptons 
and their antiparticles with the properties assumed by the {\it standard model} (with the right 
handed neutrinos in addition), connecting handedness with the weak and hyper charges. 

In this theory the hyper charge $Y$ appears as a sum of the two quantum numbers: $\tau^{4}$ 
(a "fermion charge" quantum number) and $\tau^{23}$ ($\vec{\tau}^{2}$ is the second $SU(2)$ 
charge). This is the case also for $SO(10)$.
 Taking this into account and the fact that one Weyl representation includes quarks and 
leptons and anti-quarks and anti-leptons and has the traces of all subgroups
 equal to zero, makes a simple explanation for the traceless products of all contributions
to the triangle anomalies of the {\it standard model}, with the $U(1) \times U(1) \times U(1)$ 
included. 

It should also be pointed out that embedding the {\it standard model} groups into $SO(13,1)$ makes 
the weak and hyper charges of particles and anti-particles connected with their handedness, if first 
breaking $SO(13,1)$ to $SO(7,1) \times$ $SU(3)$ $\times U(1)$ and then further to $SO(3,1)
\times SU(2)$  $\times SU(3)$ $\times U(1)$.

In the {\it spin-charge-family} theory the $SO(10)$ states can be trivially analyzed 
(Subsects.~\ref{so10},\ref{explicit}), if the "technique",  which represents spinors as products of 
$\frac{d}{2}$ nilpotents and projectors, is used. By embedding the $SO(10)$ group into the 
$SO(13,1)$ group, it becomes transparent that   the $SO(10)$ unifying theories must
connect "by hand" the spin and the handedness of $SO(3,1)$  with the charges of  $SO(10)$.

%
%(The $SO(10)$ unifying theories do not do more than the {\it standard model} with respect to 
%connecting
%charges with the handedness of particles and antiparticles.) 

%In the {\it spin-charge-family} theory  all the vector and the scalar gauge fields origin in gravity, 
%Ref.~\cite{EPJC2017} (and the references therein). The gravitational origin of these vector gauge 
%fields can be neglected as long as they are weak in comparison with fields which make ($d-4$) space 
%curled and correspondingly we do not need to know how to quantize gravity. It is assumed in this 
%paper that all the vector gauge fields are quantized as $U(1)$ and $SU(n)$ vector gauge fields. 
%Yet the fact that embedding the {\it standard model} groups into $SO(13,1)$  might be the right way 
%to explain the "miraculous" anomaly cancellation of the {\it standard model}.
%% Mention Holger Norma Bled proceeding contribution of the cut off possibility.

Let us conclude this section by pointing out that embedding the {\it standard model} groups $U(1)$,
$SU(2)$ and $SU(3)$ into $SO(13,1)$ might help to understand all  the assumptions of the 
{\it Standard model}: about the origin of spins and charges of fermions, the origin of vector gauge 
fields and also about the origin of scalar gauge fields and families of fermions, if both kinds of 
existing Clifford algebra objects are taken into account.
%again that the  {\it spin-charge-family} theory offers --- 
%starting  in 
%$d=(13 +1)$ with the simple action for massless spinors, carrying (only) both kinds of  
%spins, and for gravity fields, represented by vielbeins and the two kinds of the spin connection fields,
%all massless --- in $d=(3+1)$ the explanation for the origin of all the vector  gauge fields, for the 
%Higgs's scalar and the Yukawa couplings, for the appearance of families, for the appearance of 
%the dark matter, for the appearance of the matter-antimatter asymmetry, and other phenomena, 
%making several predictions for the future 
%measurements~\cite{IARD2016}, explaining also the "miraculous" cancellation of the triangle 
%anomaly of the {\it standard model}. 
%

%
\appendix

\section{Short presentation of the {\it spin-charge-family} theory}
\label{SCFT}

The {\it spin-charge-family} theory~\cite{EPJC2017,JMP2015,norma2014MatterAntimatter,%
NBled2013,
NBled2012,IARD2016,pikanorma,portoroz03,norma92,norma93,norma94,norma95,gmdn07,%
gn,gn2013,gn2015,NPLB,N2014scalarprop} assumes a simple action 
(Eq.~(\ref{wholeaction})) in an even dimensional space ($d=2n$, $d>5$), $d$ is chosen to be 
 $(13+1)$, what makes the action to manifest in $d=(3+1)$ in the low energy regime all the 
observed degrees of freedom, explaining all 
the assumptions of the {\it standard model}, as well as other observed phenomena.
Fermions interact with the vielbeins $f^{\alpha}{}_{a}$ and the two kinds of the spin-connection 
fields - $\omega_{ab \alpha}$ and $\tilde{\omega}_{ab \alpha}$ -
the  gauge fields of $S^{ab} = \,\frac{i}{4} (\gamma^a\, \gamma^b
- \gamma^b\, \gamma^a)\,$ and $\tilde{S}^{ab} = \,\frac{i}{4} (\tilde{\gamma}^a\, 
\tilde{\gamma}^b - \tilde{\gamma}^b\, \tilde{\gamma}^a)$, respectively. The action
\begin{eqnarray}
{\cal A}\,  &=& \int \; d^dx \; E\;\frac{1}{2}\, (\bar{\psi} \, \gamma^a p_{0a} \psi) + h.c. +
%{\mathcal L}_{f} +  
\nonumber\\  
               & & \int \; d^dx \; E\; (\alpha \,R + \tilde{\alpha} \, \tilde{R})\,,%\nonumber\\
               %\end{eqnarray}
%
%\begin{eqnarray}
%{\mathcal L}_f &=& \frac{1}{2}\, (\bar{\psi} \, \gamma^a p_{0a} \psi) + h.c., 
%\nonumber\\
%p_{0a }        &=& f^{\alpha}{}_a p_{0\alpha} + \frac{1}{2E}\,
% \{ p_{\alpha}, E f^{\alpha}{}_a\}_-, 
%\nonumber\\  
%   p_{0\alpha} &=&  p_{\alpha}  - 
%                    \frac{1}{2}  S^{ab} \omega_{ab \alpha} - 
%                    \frac{1}{2}  \tilde{S}^{ab}   \tilde{\omega}_{ab \alpha},                   
%\nonumber\\ 
%R              &=&  \frac{1}{2} \, \{ f^{\alpha [ a} f^{\beta b ]} \;(\omega_{a b \alpha, \beta} 
%- \omega_{c a \alpha}\,\omega^{c}{}_{b \beta}) \} + h.c. \;, 
%\nonumber\\
%\tilde{R}      &=&  \frac{1}{2} \, \{ f^{\alpha [ a} f^{\beta b ]} \
%;(\tilde{\omega}_{a b \alpha,\beta} - 
%\tilde{\omega}_{c a \alpha} \,\tilde{\omega}^{c}{}_{b \beta})\} + h.c.\;, 
\label{wholeaction}
\end{eqnarray}
in which $p_{0a } = f^{\alpha}{}_a\, p_{0\alpha} + \frac{1}{2E}\, \{ p_{\alpha},
E f^{\alpha}{}_a\}_- $, 
$ p_{0\alpha} =  p_{\alpha}  - \frac{1}{2} \, S^{ab}\, \omega_{ab \alpha} - 
                    \frac{1}{2} \,  \tilde{S}^{ab} \,  \tilde{\omega}_{ab \alpha} $,   and                   
$$R =  \frac{1}{2} \, \{ f^{\alpha [ a} f^{\beta b ]} \;(\omega_{a b \alpha, \beta} 
- \omega_{c a \alpha}\,\omega^{c}{}_{b \beta}) \} + h.c., $$  
$$\tilde{R}  =  \frac{1}{2} \, \{ f^{\alpha [ a} f^{\beta b ]} \;(\tilde{\omega}_{a b \alpha,\beta} - 
\tilde{\omega}_{c a \alpha} \,\tilde{\omega}^{c}{}_{b \beta})\} + h.c.$$~\footnote{Whenever 
two indexes are equal the summation over these two is meant.}, 
 introduces two kinds of the Clifford algebra objects, $\gamma^a$ and
 $\tilde{\gamma}^a$,
%
%\begin{eqnarray}
%\label{twoclifford}
$\{\gamma^a, \gamma^b\}_{+}= 2 \eta^{ab} = 
\{\tilde{\gamma}^a, \tilde{\gamma}^b\}_{+}$.
%\end{eqnarray}
% 
$f^{\alpha}{}_{a}$ are vielbeins inverted to $e^{a}{}_{\alpha}$, Latin letters ($a,b,..$) denote
flat indices, Greek letters ($\alpha,\beta,..$) are Einstein indices,  $(m,n,..)$ and $(\mu,\nu,..)$ 
denote the corresponding indices in ($0,1,2,3$), $(s,t,..)$ and $(\sigma,\tau,..)$  denote the
corresponding indices in $d\ge5$:
\begin{eqnarray}
\label{vielfe}
e^{a}{}_{\alpha}f^{\beta}{}_{a} &=&\delta^{\beta}_{\alpha}\,, \quad
e^{a}{}_{\alpha}f^{\alpha}{}_{b}= \delta^{a}_{b}\,,
\end{eqnarray}
$E =\det(e^{a}{}_{\alpha})$.

The action ${\cal A} $ offers the explanation for the origin and all the properties of the observed 
fermions (of the family members and families), of the observed vector gauge fields, of the Higgs's
scalar and of the Yukawa couplings, explaining the origin of the matter-antimatter asymmetry,  the 
appearance of the dark matter and predicts new scalars and the $4^{th}$  family  
to the known three, %and a new gauge field 
to be observed at the LHC~(\cite{norma2014MatterAntimatter,IARD2016} and the 
references therein). 

The {\it standard model} groups of spins and charges are the subgroups of the $SO(13,1)$ group 
with the generator of the infinitesimal transformations expressible with $S^{ab}$ for the spin 
\begin{eqnarray}
\label{so1+3}
\vec{N}_{\pm}(= \vec{N}_{(L,R)}): &=& \,\frac{1}{2} (S^{23}\pm i S^{01},S^{31}\pm i S^{02}, 
S^{12}\pm i S^{03} )\,,%\,,\nonumber\\
%\vec{\tilde{N}}_{\pm}(=\vec{\tilde{N}}_{(L,R)}):&=& \,\frac{1}{2} (\tilde{S}^{23}\pm i \tilde{S}^{01},
%\tilde{S}^{31}\pm i \tilde{S}^{02}, \tilde{S}^{12}\pm i \tilde{S}^{03} )\,,
\end{eqnarray}
for the weak charge, $SU(2)_{I}$, and the second $SU(2)_{II}$, these two groups are the invariant 
subgroups of $SO(4)$,
 \begin{eqnarray}
 \label{so42}
 \vec{\tau}^{1}:&=&\frac{1}{2} (S^{58}-  S^{67}, \,S^{57} + S^{68}, \,S^{56}-  S^{78} )\,,
\nonumber\\
 \vec{\tau}^{2}:&=& \frac{1}{2} (S^{58}+  S^{67}, \,S^{57} - S^{68}, \,S^{56}+  S^{78} )\,,
%\,\;\nonumber\\
 %\vec{\tilde{\tau}}^{1}:&=&\frac{1}{2} (\tilde{S}^{58}-  \tilde{S}^{67}, \,\tilde{S}^{57} + 
 %\tilde{S}^{68}, \,\tilde{S}^{56}-  \tilde{S}^{78} )\,,\;\;
 %\vec{\tilde{\tau}}^{2}:=\frac{1}{2} (\tilde{S}^{58}+  \tilde{S}^{67}, \,\tilde{S}^{57} - 
 %\tilde{S}^{68}, \,\tilde{S}^{56}+  \tilde{S}^{78} ),\,\,\;\;
 \end{eqnarray}
for the colour charge  $SU(3)$ and for the "fermion charge" $U(1)_{II}$, these two groups are
subgroups of $SO(6)$,
 \begin{eqnarray}
 \label{so64}
 \vec{\tau}^{3}: = &&\frac{1}{2} \,\{  S^{9\;12} - S^{10\;11} \,,
  S^{9\;11} + S^{10\;12} ,\, S^{9\;10} - S^{11\;12} ,\nonumber\\
 && S^{9\;14} -  S^{10\;13} ,\,  S^{9\;13} + S^{10\;14} \,,
  S^{11\;14} -  S^{12\;13}\,,\nonumber\\
 && S^{11\;13} +  S^{12\;14} ,\, 
 \frac{1}{\sqrt{3}} ( S^{9\;10} + S^{11\;12} - 
 2 S^{13\;14})\}\,,\nonumber\\
 \tau^{4}: = &&-\frac{1}{3}(S^{9\;10} + S^{11\;12} + S^{13\;14})\,,%\;\;\nonumber\\
 %\tilde{\tau}^{4}: = &&-\frac{1}{3}(\tilde{S}^{9\;10} + \tilde{S}^{11\;12} + \tilde{S}^{13\;14})\,,
 \end{eqnarray}
while the hyper charge $Y=\tau^{23} + \tau^{4}$. The breaks of the symmetries, manifesting
in Eqs.~(\ref{so1+3}, \ref{so42}, \ref{so64}), are in the {\it spin-charge-family} theory caused by
the condensate and the constant values of the scalar fields carrying the space index $(7,8)$ 
(Refs.~\cite{JMP2015,IARD2016} and the references therein). The space breaks first to $SO(7,1)$
$\times SU(3) \times U(1)_{II}$ and then further to $SO(3,1)\times SU(2)_{I} \times U(1)_{I}$
$\times SU(3) \times U(1)_{II}$, what explains the connections between the weak and the hyper 
charges and the handedness of spinors.

The equivalent expressions for the family charges, expressed by $\tilde{S}^{ab}$ follow if in 
Eqs.~(\ref{so1+3} - \ref{so64}) $S^{ab}$ are replaced by $\tilde{S}^{ab}$. 

\subsection{Evaluation of some properties of spinor states explicitly}
\label{explicit}

We demonstrate in this subsection on few examples how (elegantly) can properties of spinor states be
extracted from the expressions with nilpotents and projectors in Table~\ref{Table so13+1.}. We 
comment also on the $SO(10)$ unification scheme, comparing this scheme with the $SO(13,1)$ 
of the {\it spin-charge-family} theory.

\vspace{2mm}

{\it Example 1.}: Let us calculate properties of  the two states: $ \stackrel{03}{(+i)}\stackrel{12}{(+)}|
\stackrel{56}{(+)}\stackrel{78}{(+)} ||\stackrel{9 \;10}{(+)}\stackrel{11\;12}{(-)}
\stackrel{13\;14}{(-)} |\psi_{0} \rangle$ and $ \stackrel{03}{[-i]}\stackrel{12}{[-]}|
\stackrel{56}{(+)}\stackrel{78}{(+)} ||\stackrel{9 \;10}{(+)}\stackrel{11\;12}{(-)}
\stackrel{13\;14}{(-)} |\psi_{0} \rangle$, 
appearing in the first and second line of Table~\ref{Table so13+1.} and representing $u_{R}^{c1}$ 
with the spin up and down, respectively.

From Eq.~(\ref{hand}) one calculates the handedness for the whole one Weyl representation (with 
$64$ states): $\Gamma^{(d=14)}=$  $i^{d-1}\cdot 2^{\frac{d}{2}} \cdot S^{0 3} \cdot S^{12}\cdots 
%$(-2i)^7  S^{03}  S^{12}  S^{56}  S^{78}  S^{9\,10}  S^{11\,12}  
S^{13\,14}  $. Applying this operator on the first state of Table~\ref{Table so13+1.} by using
Eq.~(\ref{grapheigen}),  gives  as the eigenvalue $=i^{13} 2^{7} \frac{i}{2}  (\frac{1}{2})^6 
=-1$ (since the operator $S^{03}$ on $ \stackrel{03}{(+i)}$ gives the eigenvalue $\frac{k}{2}= 
\frac{i}{2} $, the next four operators have the eigenvalues on the corresponding nilpotents 
%Eq.~(\ref{choicecartan}), 
equal to $ \frac{1}{2}$, and the last two  $- \frac{1}{2}$). One easily sees that the second state
has also the same handedness $\Gamma^{(14)}$ as the whole representation. 

In an equivalent way we calculate the  handedness $\Gamma^{(3,1)}$ of  these two states:   
%in $d=(3+1)$:  
Applying $\Gamma^{(3,1)} =  i^3\cdot  2^2\cdot S^{03} S^{12}$ on any of these two   
states gives $1$ ($i^3 2^2 \cdot  \frac{i}{2}  \frac{1}{2} $ and $i^3 2^2 \cdot (-\frac{i}{2}) 
(- \frac{1}{2}) $, respectively) --- the right handedness. 

The weak charge operator $\tau^{13}=\frac{1}{2}\, (S^{56}-S^{78})$, Eq.~(\ref{so42}), 
applied on any of these two states, gives $0$: $\frac{1}{2}\, (\frac{1}{2}-\frac{1}{2})$, 
correspondingly 
$\tau^{23}$ $=\frac{1}{2}\, (S^{56} + S^{78})$, applied on any of these two states,  gives
$\frac{1}{2}$, while the "fermion charge" operator $\tau^4 = - \frac{1}{3} ( S^{9\,10} + 
S^{11\,12} + S^{13\,14} )$, Eq.~(\ref{so64}), gives %applied on any of these two states, 
$- \frac{1}{3} ( \frac{1}{2} - \frac{1}{2} - \frac{1}{2}) = \frac{1}{6}$. Correspondingly  the
hyper charge operator $Y=\tau^{23} + \tau^{4}$ gives for these two states $Y= \frac{2}{3} $
(what the {\it standard model} assumes for $u_{R}$). 

One finds for the colour charge of these two states (all the operators $S^{ab}$ appearing 
in the expressions for the colour charge are diagonal),  when applying 
$\tau^{3 3}$ $(=\frac{1}{2} (S^{9\,10} - S^{11\,12}))$  and
$\tau^{3 8}$ $(=\frac{1}{\sqrt{3}} (S^{9\,10} + S^{11\,12} - 2 S^{13\,14}))$ on these states,
the eigenvalues $ (1/2\,,1/(2\sqrt{3})$, respectively.

These two states differ in the spin, carrying ($\frac{1}{2},-\frac{1}{2}$), respectively. All the 
states of this octet --- $SO(7,1)$ --- have the same "fermion charge" and the same colour charge.

One can calculate in the same way the properties of the left handed $u_{L}$-quarks, belonging to
this  octet: both $u_{L}$-quark, with the eigenvalues of $S^{12}$ equal to $\pm \frac{1}{2}$,
carry the left handedness $(-1)$, the weak charge $\tau^{13}=\frac{1}{2}$, the charge  
$\tau^{23}=0$ and the hyper charge $Y=\frac{1}{6}$~\footnote{We denote the operators and 
their eigenvalues with the same mark.}.  
The left and right handed members of this octet carry obviously the handedness and the spin in
the chiral representation. It is not difficult to write down  matrix representation for all the 
{\it standard model} operators for spins and charges used in the text books.

 Let us point out that the octet $SO(7,1)$ manifests how the spin and the weak and hyper charges 
are correlated.

In the same way one calculates the  properties also of the left and right haded $d_{R,L}$-quarks, 
belonging to the same octet.

All the octets, the $SO(7,1)$ ($\Gamma^{(7,1)}= 1$) part of the $SO(13,1)$ spinor representation, 
are the same for either 
quarks of the additional two colours (quarks states appear in Table~\ref{Table so13+1.} from the 
first to the $24^{th}$ line), or for the colourless leptons  (leptons appear in Table~\ref{Table so13+1.}
from the $25^{th}$ line to the $32^{nd}$ line). 

Leptons distinguish  from quarks in the part represented by nilpotents and projectors, which is determined
by the Cartan subalgebra of ($S^{9\,10}, S^{11\,12}, S^{13\,14}$). Taking into account 
Eq.~(\ref{grapheigen}) one calculates that ($\tau^{3 3}$, $\tau^{3 8}$) is for the colour (chargeless) 
part of lepton states  $(\nu_{R,L}, e_{R,L})$ ---  
($\cdots| \cdots||\stackrel{9 \;10}{(+)}\;\;\stackrel{11\;12}{[+]}\;\;\stackrel{13\;14}{[+]} $) ---
equal to $(0, 0)$, while the "fermion charge" $\tau^{4}$ is for these states equal to $-\frac{1}{2}$
(just as the {\it standard model} requires).

\vspace{2mm}

{\it Example 2.}: To the same fundamental representation of $SO(13,1)$ belong besides quarks and
leptons also anti-quarks (presented in Table~\ref{Table so13+1.} from the $33^{rd}$ line to the 
$56^{th}$ line)
and anti-leptons (presented in Table~\ref{Table so13+1.} as the last octet, lines from $57$  to  
$64$). Again,
all the anti-octets, the ${\bar SO}(7,1)$ ($\Gamma^{(7,1)}= - 1$) part of the $SO(13,1)$ 
fundamental representation, are the 
same either for anti-quarks or for anti-leptons, anti-quarks appearing in the three anti-colours, 
anti-leptons in the anti-colourless state.  

Let us add that all the anti-spinor states are reachable from the spinor states (and opposite) 
 by applying on the particle states the operator~\cite{discretesym} 
$\mathbb{C}_{{\cal N}}  {\cal P}_{{\cal N}}$~\footnote{Discrete symmetries in $d=(3+1)$ follow
 from the corresponding definition of these symmetries in $d$-
dimensional space~\cite{discretesym,TDN}.  
The operator is defined as: $\mathbb{C}_{{\cal N}}  {\cal P}_{{\cal N}} $ $=\gamma^0
 \prod_{\Im \gamma^s, s=5}^{d} \gamma^s\,\,I_{\vec{x}_3} \,I_{x^6,x^8,\dots,x^{d}}\,$,
 where $\gamma^0$ and $\gamma^1$ are real, 
$\gamma^2$ imaginary, $\gamma^3$ real, $\gamma^5$ imaginary, $\gamma^6$ real, alternating 
imaginary and real up to $\gamma^d$, which is in even dimensional spaces real.  
$\gamma^a$'s appear in the ascending order.
Operators $I$ operate as follows: $\quad I_{x^0} x^0 = -x^0\,$; $
I_{x} x^a =- x^a\,$; $  I_{x^0} x^a = (-x^0,\vec{x})\,$; $ I_{\vec{x}} \vec{x} = -\vec{x}
\,$; $I_{\vec{x}_{3}} x^a = (x^0, -x^1,-x^2,-x^3,x^5, x^6,\dots, x^d)\,$; 
$I_{x^5,x^7,\dots,x^{d-1}}$ $(x^0,x^1,x^2,x^3,x^5,x^6,x^7,x^8,
\dots,x^{d-1},x^d)$ $=(x^0,x^1,x^2,x^3,-x^5,x^6,-x^7,\dots,-x^{d-1},x^d)$;
 $I_{x^6,x^8,\dots,x^d}$ 
$(x^0,x^1,x^2,x^3,x^5,x^6,x^7,x^8,\dots,$ $x^{d-1},x^d)$
$=(x^0,x^1,x^2,x^3,x^5,-x^6,x^7,-x^8,\dots$ $,x^{d-1},-x^d)$, $d=2n$.}.
The part of this operator,
 which operates on only the spinor part of the state (presented in Table~\ref{Table so13+1.})
is $\mathbb{C}_{{\cal N}}  {\cal P}_{{\cal N}}|_{spinor}$ 
$=\gamma^0  \prod_{\Im \gamma^s, s=5}^{d} \gamma^s$. 
Taking into account 
Eq.~(\ref{snmb:gammatildegamma}) and this operator  one finds that
 $\mathbb{C}_{{\cal N}}  {\cal P}_{{\cal N}}|_{spinor}$
transforms $u^{c1}_{R}$ from the first line of Table~\ref{Table so13+1.} into 
${\bar u}^{\bar{c1}}_{L}$ from the $35^{th}$ line of the same table. When the operator 
$\mathbb{C}_{{\cal N}}  {\cal P}_{{\cal N}}|_{spinor}$ applies on $\nu_{R}$ (the 
$25^{th}$ line of the same table) --- with the colour chargeless part  equal to 
$\cdots||\stackrel{9 \;10}{(+)}\;\;\stackrel{11\;12}{[+]}\;\;\stackrel{13\;14}{[+]} $) ---   
transforms  $\nu_{R}$ into 
${\bar \nu}_{L}$  (the $59^{th}$ line of the table) --- with the colour anti-chargeless part equal to
$\cdots||\stackrel{9 \;10}{[-]}\;\;\stackrel{11\;12}{(-)}\;\;\stackrel{13\;14}{(-)} $.

All the states of one fundamental representation are reachable from any state of the representation 
by the generators of the infinitesimal transformations $S^{ab}$. 

\vspace{2mm}

{\it Example 3.}: Applying the operators $\tilde{S}^{ab}$ (they commute with $S^{ab}$, 
Eq.~(\ref{sabtildesab})), which does not belong to the chosen Cartan subalgebra,  
Eq.~(\ref{choicecartan}), on a spinor state of % the fundamental representation of   
Table~\ref{Table so13+1.},  the particular state transforms to the same family member state of 
another family, which is orthogonal to the starting state. 

Let us apply, for example, $\tilde{S}^{57}$ on $\nu_{R}$ (the $25^{th}$ line of 
Table~\ref{Table so13+1.}, represented in Table~\ref{Table III.} as  the $\nu_{R8}$ state
in the eighth line on the right hand side of the table). Since 
$\tilde{S}^{57}= \frac{i}{2}\tilde{\gamma}^5 \tilde{\gamma}^7$, one obtains by taking into 
account Eq.~(\ref{snmb:gammatildegamma})  the $\nu_{R6}$ state belonging to another 
family, presented in the sixth line of the right part of Table~\ref{Table III.}.

\vspace{2mm}

{\it Example 4.}: Let us insert $SO(10)\times SO(3,1)$ into $SO(13,1)$ to recognize
 what does $SO(13,1)$ group offer more than  $SO(10)$.   

$SO(10)$ unifying models pay attention on only the charge part of particles, unifying all the charges, 
while they connect $SO(10)$ to $SO(3,1)$ "by hand". These models also do not have the 
explanation for the families, for the origin of the vector gauge fields and for the Higg's scalar and 
the Yukawa couplings, at least not in an unique way.

Let us correspondingly pay attention to only the $SO(10)$ part of the $SO(13,1)$
representation of Table~\ref{Table so13+1.}. The fundamental representation of $SO(10)$ has 
$2^{\frac{d}{2}-1}= 16$ states, while the $SO(13,1)$ has $4$ times more states, as expected 
(no spin parts and handedness of states are included in $SO(10)$, what brings a factor of $4$).

Let us start with the $u_{R}^{c1}$-quark state  ($1^{st}$ line of  Table~\ref{Table so13+1.}), 
neglecting the spin part (the first two nilpotents $ \stackrel{03}{(+i)}\,\stackrel{12}{(+)}|$ ) of the
state, requiring ("by hand") that what is left belongs to 
$u_{R}^{c1}$-quark state. The operator $S^{57}=\frac{i}{2} \gamma^5 \gamma^7$ 
(let Eq.~(\ref{snmb:gammatildegamma}) be used) transforms $u_{R}^{c1}$ into $d_{R}^{c1}$ 
(the $3^{rd}$ state of the table)  keeping the spin and handedness, which was assumed for 
$u_{R}^{c1}$. 
%(The operator $S^{57}$  does not influence spin or handedness).
The operator $S^{9\,11}=\frac{i}{2} \gamma^9 \gamma^{11}$, 
Eq.~(\ref{snmb:gammatildegamma}), transforms $u_{R}^{c1}$ into
$u_{R}^{c2}$ (the $9^{th}$ line) and  $d_{R}^{c1}$ into  $d_{R}^{c2}$ (the $11^{th}$ line), 
$S^{9\,13}=\frac{i}{2} \gamma^9 \gamma^{13}$ transforms $u_{R}^{c1}$ into
$u_{R}^{c3}$ (the $17^{th}$ line) and  $d_{R}^{c1}$ into  $d_{R}^{c3}$ (the $19^{th}$ line),
while the operator $S^{11\,13}=\frac{i}{2} \gamma^{11} \gamma^{13}$ transforms $u_{R}^{c1}$ 
into $\nu_{R}$ (the $25^{th}$ line) and  $d_{R}^{c1}$ into  $e_{R}$ (the $27^{th}$ line). 
The last three transformations concern only the colour and "fermion charge" part of $SO(10)$, 
keeping the weak ($SU(2)_{I}$) charge and the $SU(2)_{II}$ part of the hyper charge unchanged. 

Let us now allow the transformations from the weak and the $SU(2)_{II}$ part of the hyper 
charges to the colour and the "fermion charge" ($U(1)_{II}$) part. The operator $S^{7\, 9}=
\frac{i}{2} \gamma^7 \gamma^{9}$ transforms $u_{R}^{c1}$ into
${\bar e}_{R}$ (the $63^{th}$ line) and  $d_{R}^{c1}$ into  ${\bar \nu}_{R}$
 (the $61^{st}$ line), 
$S^{7\, 11}=\frac{i}{2} \gamma^7 \gamma^{11}$ transforms $u_{R}^{c1}$ into
${\bar d}^{\bar c3}_{R}$ (the $53^{rd}$ line) and  $d_{R}^{c1}$ into  
${\bar u}^{{\bar c3}}_{R}$ 
(the $55^{th}$ line), and $S^{7\, 13}=\frac{i}{2} \gamma^7 \gamma^{13}$ transforms 
$u_{R}^{c1}$ into
${\bar d}^{\bar c2}_{R}$ (the $45^{th}$ line) and  $d_{R}^{c1}$ into  
${\bar u}^{{\bar c2}}_{R}$ 
(the $47^{th}$ line). The operator $S^{9\, 11}=\frac{i}{2} \gamma^9 \gamma^{11}$ 
transforms ${\bar d}^{\bar c2}_{R}$ into ${\bar d}^{\bar c1}_{R}$  (the  $37^{th}$ line) and
 ${\bar u}^{\bar c2}_{R}$ into ${\bar u}^{\bar c1}_{R}$  (the  $39^{th}$ line). 
The above $16$ states are presented in  the lower part of 
Table~\ref{Table SMandSCFTspinors.} .

One gets the right handed representation of $SO(10)$ with spin $\frac{1}{2}$, presented in the 
lower part of Table~\ref{Table SMandSCFTspinors.} by requiring the appropriate (right in this case)
handedness. (To get the right handed representation with spin down one should start with the 
right handed $u_{R}^{c1}$ with spin $-\frac{1}{2}$.)

To get the left handed representation of Table~\ref{Table SMandSCFTspinors.} (with spin up or down) 
one should start with the left handed $u_{L}^{c1}$ (with spin up or down) and repeat all the above 
procedure. The decision what is the handedness of the spinor state is made "by hand", like in the 
{\it standard model}.

\subsection{Short presentation of spinor technique~\cite{IARD2016,JMP2015,norma93,hn02,hn03}}
\label{technique}

This appendix is a short review (taken from~\cite{JMP2015}) of the technique~\cite{norma93,DKhn,%
hn02,hn03}, initiated and developed in Ref.~\cite{norma93} by one of the authors (N.S.M.B.), while  
proposing the {\it spin-charge-family} theory~\cite{EPJC2017,JMP2015,%
norma2014MatterAntimatter,NBled2013,NBled2012,IARD2016,pikanorma,portoroz03,norma92,%
norma93,norma94,norma95,gmdn07,gn,gn2013,gn2015,NPLB,N2014scalarprop}.
All the internal degrees of freedom of spinors, with family quantum numbers included, are describable 
with two kinds of the Clifford algebra objects, besides with $\gamma^a$'s, used in this theory to describe
spins and all the charges of fermions, also with  $\tilde{\gamma}^a$'s, used in this theory to describe
families of spinors:
\begin{eqnarray}
\label{gammatildegamma}
&& \{ \gamma^a, \gamma^b\}_{+} = 2\eta^{ab}\,, \quad\quad    
\{ \tilde{\gamma}^a, \tilde{\gamma}^b\}_{+}= 2\eta^{ab}\,, \quad\quad
\{ \gamma^a, \tilde{\gamma}^b\}_{+} = 0\,.
\end{eqnarray}
We assume  the ``Hermiticity'' property for $\gamma^a$'s  (and $\tilde{\gamma}^a$'s) 
%\begin{eqnarray}
$\gamma^{a\dagger} = \eta^{aa} \gamma^a$ (and $\tilde{\gamma}^{a\dagger} = 
\eta^{aa} \tilde{\gamma}^a$),
%\label{cliffher}
%\end{eqnarray}
%
in order that $\gamma^a$ (and $\tilde{\gamma}^a$) are compatible with (\ref{gammatildegamma}) 
and formally unitary, i.e. $\gamma^{a \,\dagger} \,\gamma^a=I$
 (and $\tilde{\gamma}^{a\,\dagger} \tilde{\gamma}^a=I$).
%We could make for \tilde{\gamma}^{a\dagger} =- \eta^{aa} \tilde{\gamma}^a\,, which would 
% also be acceptable as an antiunitary operator and everything would work. Use this for $d=(5+1)$.
One correspondingly finds  that $(S^{ab})^{\dagger} = \eta^{aa} \eta^{bb}S^{ab}$ (and 
 $(\tilde{S}^{ab})^{\dagger} = \eta^{aa} \eta^{bb} \tilde{S}^{ab}$). 

Spinor states are represented as products of nilpotents and projectors, formed as odd and even 
objects of $\gamma^a$'s, respectively, chosen to be the eigenstates of a Cartan subalgebra of 
the Lorentz  groups defined by $\gamma^a$'s
\begin{eqnarray}
\stackrel{ab}{(k)}:&=& 
\frac{1}{2}(\gamma^a + \frac{\eta^{aa}}{ik} \gamma^b)\,,\quad \quad
\stackrel{ab}{[k]}:=
\frac{1}{2}(1+ \frac{i}{k} \gamma^a \gamma^b)\,,%\nonumber\\
%\stackrel{+}{\circ}:&=& \frac{1}{2} (1+\Gamma)\,,\quad \quad
%\stackrel{-}{\bullet}:= \frac{1}{2}(1-\Gamma),
\label{signature}
\end{eqnarray}
where $k^2 = \eta^{aa} \eta^{bb}$. 
We further have~\cite{JMP2015}
\begin{eqnarray}
\gamma^{a}\,\stackrel{ab}{(k)}:&=& 
\frac{1}{2}(\gamma^a  \gamma^a + \frac{\eta^{aa}}{ik} \gamma^a \gamma^b)=
\eta^{aa}\,\stackrel{ab}{[-k]}\,,\quad \gamma^{a}\,\stackrel{ab}{[k]}:=
\frac{1}{2}(\gamma^a+ \frac{i}{k} \gamma^a \gamma^a \gamma^b)=\stackrel{ab}{(-k)}\,,\nonumber\\
\tilde{\gamma}^{a}\,\stackrel{ab}{(k)}:&=& 
-i \frac{1}{2}(\gamma^a + \frac{\eta^{aa}}{ik} \gamma^b) \gamma^a= -i \eta^{aa} 
 \stackrel{ab}{[k]}\,,\quad \tilde{\gamma}^{a}\,\stackrel{ab}{[k]}:=
i \frac{1}{2}(1+ \frac{ i}{k}\gamma^a \gamma^b) \gamma^a=- i \stackrel{ab}{(k)}\,,
%\stackrel{+}{\circ}:&=& \frac{1}{2} (1+\Gamma)\,,\quad \quad
%\stackrel{-}{\bullet}:= \frac{1}{2}(1-\Gamma),
\label{signature1}
\end{eqnarray}
where we assume that all the operators apply on the vacuum state  $|\psi_0\rangle$. 
We define a vacuum state $|\psi_0>$ so that one finds
%
%\begin{eqnarray}
$< \;\stackrel{ab}{(k)}^{\dagger}  \stackrel{ab}{(k)}\; > = 1\,, \quad
< \;\stackrel{ab}{[k]}^{\dagger}
 \stackrel{ab}{[k]}\; > = 1$.
%\label{graphherscal}
%\end{eqnarray}
%

We recognize %~(Eq.~\ref{graphgammaaction},\ref{gammatilde}) 
that $\gamma^a$ 
transform  $\stackrel{ab}{(k)}$ into  $\stackrel{ab}{[-k]}$, never to $\stackrel{ab}{[k]}$, 
while $\tilde{\gamma}^a$ transform  $\stackrel{ab}{(k)}$ into $\stackrel{ab}{[k]}$, never to 
$\stackrel{ab}{[-k]}$ 
\begin{eqnarray}
%\label{snmb:graphgammatilgegammaaction}
&&\gamma^a \stackrel{ab}{(k)}= \eta^{aa}\stackrel{ab}{[-k]},\; 
\gamma^b \stackrel{ab}{(k)}= -ik \stackrel{ab}{[-k]}, \; 
\gamma^a \stackrel{ab}{[k]}= \stackrel{ab}{(-k)},\; 
\gamma^b \stackrel{ab}{[k]}= -ik \eta^{aa} \stackrel{ab}{(-k)}\,,\nonumber\\
&&\tilde{\gamma^a} \stackrel{ab}{(k)} = - i\eta^{aa}\stackrel{ab}{[k]},\;
\tilde{\gamma^b} \stackrel{ab}{(k)} =  - k \stackrel{ab}{[k]}, \;
\tilde{\gamma^a} \stackrel{ab}{[k]} =  \;\;i\stackrel{ab}{(k)},\; 
\tilde{\gamma^b} \stackrel{ab}{[k]} =  -k \eta^{aa} \stackrel{ab}{(k)}\,. 
\label{snmb:gammatildegamma}
\end{eqnarray}

The Clifford algebra objects $S^{ab}$ and $\tilde{S}^{ab}$ close the algebra of the Lorentz 
group 
\begin{eqnarray}
\label{sabtildesab}
S^{ab}: &=& (i/4) (\gamma^a \gamma^b - \gamma^b \gamma^a)\,, \nonumber\\
\tilde{S}^{ab}: &=& (i/4) (\tilde{\gamma}^a \tilde{\gamma}^b 
- \tilde{\gamma}^b \tilde{\gamma}^a)\,,
\end{eqnarray}
$ \{S^{ab}, \tilde{S}^{cd}\}_{-}= 0\,$, %\nonumber\\
$\{S^{ab},S^{cd}\}_{-} = $ $ i(\eta^{ad} S^{bc} + \eta^{bc} S^{ad} - \eta^{ac} S^{bd} - \eta^{bd} S^{ac})\,$,
$\{\tilde{S}^{ab},\tilde{S}^{cd}\}_{-} $ $= i(\eta^{ad} \tilde{S}^{bc} + \eta^{bc} \tilde{S}^{ad} 
- \eta^{ac} \tilde{S}^{bd} - \eta^{bd} \tilde{S}^{ac})\,$.

One can easily check that the nilpotent $\stackrel{ab}{(k)}$ and the projector $\stackrel{ab}{[k]}$ 
are "eigenstates" of $S^{ab}$ and $\tilde{S}^{ab}$ 
\begin{eqnarray}
        &&S^{ab}\, \stackrel{ab}{(k)}= \frac{1}{2}\,k\, \stackrel{ab}{(k)}\,,\quad \quad 
        S^{ab}\, \stackrel{ab}{[k]}= \;\;\frac{1}{2}\,k \,\stackrel{ab}{[k]}\,,\nonumber\\
&&\tilde{S}^{ab}\, \stackrel{ab}{(k)}= \frac{1}{2}\,k \,\stackrel{ab}{(k)}\,,\quad \quad 
\tilde{S}^{ab}\, \stackrel{ab}{[k]}=-\frac{1}{2}\,k \,\stackrel{ab}{[k]}\,,
\label{grapheigen}
\end{eqnarray}
where the vacuum state $|\psi_0\rangle$ is meant to stay on the right hand sides of projectors 
and nilpotents. This means that multiplication of nilpotents $\stackrel{ab}{(k)}$ 
and projectors $\stackrel{ab}{[k]}$ by $S^{ab}$  get the same objects back multiplied 
by the constant $\frac{1}{2}k$, while $\tilde{S}^{ab}$ multiply $\stackrel{ab}{(k)}$ by 
$\frac{k}{2}$ and $\stackrel{ab}{[k]}$ by $(-\frac{k}{2})$ (rather than by $\frac{k}{2}$). 
This also means that when 
$\stackrel{ab}{(k)}$ and $\stackrel{ab}{[k]}$ act from the left hand side on  a
vacuum state $|\psi_0\rangle$ the obtained states are the eigenvectors of $S^{ab}$.
  
The technique can be used to construct a spinor basis for any dimension $d$ and any signature in an 
easy and transparent way. Equipped with nilpotents and projectors of Eq.~(\ref{signature}),  
the technique offers an elegant way to see all the quantum numbers of states with respect to the
two Lorentz groups, as well as transformation properties of the states under the application of any 
Clifford algebra object.

Recognizing from Eq.~(\ref{sabtildesab})  that the two Clifford algebra objects ($S^{ab}, S^{cd}$)
with all indexes different commute (and equivalently for ($\tilde{S}^{ab},\tilde{S}^{cd}$)), we  
select  the  Cartan subalgebra of the algebra of the 
two groups, which  form  equivalent representations with respect to one another 
\begin{eqnarray}
S^{03}, S^{12}, S^{56}, \cdots, S^{d-1\; d}, \quad {\rm if } \quad d &=& 2n\ge 4,
%\nonumber\\
%S^{03}, S^{12}, \cdots, S^{d-2 \;d-1}, \quad {\rm if } \quad d &=& (2n +1) >4\,,
\nonumber\\
\tilde{S}^{03}, \tilde{S}^{12}, \tilde{S}^{56}, \cdots, \tilde{S}^{d-1\; d}, 
\quad {\rm if } \quad d &=& 2n\ge 4\,.
%\nonumber\\
%\tilde{S}^{03}, \tilde{S}^{12}, \cdots, \tilde{S}^{d-2 \;d-1}, \quad {\rm if } \quad
% d &=& (2n +1) >4\,.
\label{choicecartan}
\end{eqnarray}

The choice of  the Cartan subalgebra in $d <4$ is straightforward.
It is  useful  to define one of the Casimirs of the Lorentz group ---  
the  handedness $\Gamma$ ($\{\Gamma, S^{ab}\}_- =0$) (as well as $\tilde{\Gamma}$)
in any $d=2n$ 
\begin{eqnarray}
\Gamma^{(d)} :&=&(i)^{d/2}\; \;\;\;\;\;\prod_a \quad (\sqrt{\eta^{aa}} \gamma^a), \quad {\rm if } \quad d = 2n, 
\nonumber\\
\tilde{\Gamma}^{(d)} :&=& (i)^{(d-1)/2}\; \prod_a \quad (\sqrt{\eta^{aa}} \tilde{\gamma}^a), 
\quad {\rm if } \quad d = 2n\,.
\label{hand}
\end{eqnarray}
%
%(One proceeds equivalently for $\tilde{\Gamma}^{(d)} $, substituting $\gamma^a$'s by 
%$\tilde{\gamma}^a$'s.)
 We understand the product of $\gamma^a$'s in the ascending order with 
respect to the index $a$: $\gamma^0 \gamma^1\cdots \gamma^d$. 
It follows from the Hermiticity properties of $\gamma^a $ %Eq.(\ref{cliffher})
for any choice of the signature $\eta^{aa}$ that $\Gamma^{\dagger}= \Gamma,\;
\Gamma^2 = I. ($ Equivalent relations are valid for $\tilde{\Gamma}$.)
We also find that for $d$ even the handedness  anticommutes with the Clifford algebra objects 
$\gamma^a$ ($\{\gamma^a, \Gamma \}_+ = 0$) (while for $d$ odd it commutes with  
$\gamma^a$ ($\{\gamma^a, \Gamma \}_- = 0$)). 

Taking into account the above equations it is easy to find a Weyl spinor irreducible representation
for $d$-dimensional space, with $d$ even or odd~\footnote{For $d$ odd
 the basic states are products of $(d-1)/2$ nilpotents and a factor $(1\pm \Gamma)$.}. 
For $d$ even we simply make a starting state as a product of $d/2$, let us say, only nilpotents 
$\stackrel{ab}{(k)}$, one for each $S^{ab}$ of the Cartan subalgebra elements
(Eqs.(\ref{choicecartan}, \ref{sabtildesab})), applying it on an (unimportant) vacuum 
state. 
Then the generators $S^{ab}$, which do not belong to the Cartan subalgebra, being applied on 
the starting state from the left hand side,  generate all the members of one Weyl spinor.  
\begin{eqnarray}
\stackrel{0d}{(k_{0d})} \stackrel{12}{(k_{12})} \stackrel{35}{(k_{35})}\cdots 
\stackrel{d-1\;d-2}{(k_{d-1\;d-2})}
|\psi_0 \,>\nonumber\\
\stackrel{0d}{[-k_{0d}]} \stackrel{12}{[-k_{12}]} \stackrel{35}{(k_{35})}\cdots 
\stackrel{d-1\;d-2}{(k_{d-1\;d-2})}
|\psi_0 \,>\nonumber\\
\stackrel{0d}{[-k_{0d}]} \stackrel{12}{(k_{12})} \stackrel{35}{[-k_{35}]}\cdots 
\stackrel{d-1\;d-2}{(k_{d-1\;d-2})}
|\psi_0 \,>\nonumber\\
\vdots \nonumber\\
\stackrel{0d}{[-k_{0d}]} \stackrel{12}{(k_{12})} \stackrel{35}{(k_{35})}\cdots 
\stackrel{d-1\;d-2}{[-k_{d-1\;d-2}]}
|\psi_0 \,>\nonumber\\
\stackrel{od}{(k_{0d})} \stackrel{12}{[-k_{12}]} \stackrel{35}{[-k_{35}]}\cdots 
\stackrel{d-1\;d-2}{(k_{d-1\;d-2})}
|\psi_0\,> \nonumber\\
\vdots 
\label{graphicd}
\end{eqnarray}
All the states have the same handedness $\Gamma $, since $\{ \Gamma, S^{ab}\}_{-} = 0$. 
States, belonging to one multiplet  with respect to the group $SO(q,d-q)$, that is to one
irreducible representation of spinors (one Weyl spinor), can have any phase. We could make a choice
of the simplest one, taking all  phases equal to one. (In order to have the usual transformation
properties for spinors under the rotation of spin and under  
${\cal C}_{{\cal N}}$ ${\cal P}_{{\cal N}}$, some of the states must be multiplied by $(-1)$.)

The above  representation demonstrates that for $d$ even 
all the states of one irreducible Weyl representation of a definite handedness follow from a starting 
state, which is, for example, a product of nilpotents $\stackrel{ab}{(k_{ab})}$, by transforming all 
possible pairs of $\stackrel{ab}{(k_{ab})} \stackrel{mn}{(k_{mn})}$ into
 $\stackrel{ab}{[-k_{ab}]} \stackrel{mn}{[-k_{mn}]}$.
There are $S^{am}, S^{an}, S^{bm}, S^{bn}$, which do this.
The procedure gives $2^{(d/2-1)}$ states. A Clifford algebra object $\gamma^a$ being applied 
from  the left hand side, transforms  a Weyl spinor of one handedness into a Weyl spinor of the 
opposite handedness. 
% Both Weyl spinors form a Dirac spinor in $d$. 

We shall speak about left handedness when $\Gamma= -1$ and about right
handedness when $\Gamma =1$% for either $d$ even or odd
.

While $S^{ab}$, which do not belong to the Cartan subalgebra (Eq.~(\ref{choicecartan})), generate 
all the states of one representation,  $\tilde{S}^{ab}$, which do not belong to the 
Cartan subalgebra (Eq.~(\ref{choicecartan})),  generate the states of $2^{d/2-1}$ equivalent
 representations.

Making a choice of the Cartan subalgebra set (Eq.~(\ref{choicecartan})) 
of the algebra $S^{ab}$ and 
$\tilde{S}^{ab}$:   
%
%\begin{eqnarray}
($S^{03}, S^{12}, S^{56}, S^{78}, S^{9 \;10}, S^{11\;12}, S^{13\; 14}\,$), %\nonumber\\
($\tilde{S}^{03}, \tilde{S}^{12}, \tilde{S}^{56}, \tilde{S}^{78}, \tilde{S}^{9 \;10}, 
\tilde{S}^{11\;12}, \tilde{S}^{13\; 14}\,$),
%\label{cartan}
%\end{eqnarray}
%
a left handed ($\Gamma^{(13,1)} =-1$) eigenstate of all the members of the 
Cartan  subalgebra, representing a weak chargeless  $u_{R}$-quark with spin up, hyper charge ($2/3$) 
and  colour ($1/2\,,1/(2\sqrt{3})$), for example, can be written as %(Eq.(\ref{cartan})) 
\begin{eqnarray}
&& \stackrel{03}{(+i)}\stackrel{12}{(+)}|\stackrel{56}{(+)}\stackrel{78}{(+)}
||\stackrel{9 \;10}{(+)}\stackrel{11\;12}{(-)}\stackrel{13\;14}{(-)} |\psi_{0} \rangle = \nonumber\\
&&\frac{1}{2^7} 
(\gamma^0 -\gamma^3)(\gamma^1 +i \gamma^2)| (\gamma^5 + i\gamma^6)(\gamma^7 +i \gamma^8)||
\nonumber\\
&& (\gamma^9 +i\gamma^{10})(\gamma^{11} -i \gamma^{12})(\gamma^{13}-i\gamma^{14})
|\psi_{0} \rangle \,.
\label{start}
\end{eqnarray}
This state is an eigenstate of all $S^{ab}$ and $\tilde{S}^{ab}$ which are members of the Cartan 
subalgebra (Eq.~(\ref{choicecartan})). 

The operators $ \tilde{S}^{ab}$, which do not belong to the Cartan subalgebra
 (Eq.~(\ref{choicecartan})), generate families from the starting $u_R$ quark, transforming the 
$u_R$ quark from Eq.~(\ref{start}) to the $u_R$ of another family,  keeping all of the properties 
with respect to $S^{ab}$ unchanged.
In particular, $\tilde{S}^{01}$ applied on a right handed $u_R$-quark 
%, weak chargeless,  with spin up,hypercharge ($2/3$) and the colour charge ($1/2\,,1/(2\sqrt{3})$) 
from Eq.~(\ref{start}) generates a 
state which is again  a right handed $u_{R}$-quark,  weak chargeless,  with spin up,
hyper charge ($2/3$)
and the colour charge ($1/2\,,1/(2\sqrt{3})$)
\begin{eqnarray}
\label{tildesabfam}
\tilde{S}^{01}\;
\stackrel{03}{(+i)}\stackrel{12}{(+)}| \stackrel{56}{(+)} \stackrel{78}{(+)}||
\stackrel{9 10}{(+)} \stackrel{11 12}{(-)} \stackrel{13 14}{(-)}= -\frac{i}{2}\,
&&\stackrel{03}{[\,+i]} \stackrel{12}{[\,+\,]}| \stackrel{56}{(+)} \stackrel{78}{(+)}||
\stackrel{9 10}{(+)} \stackrel{11 12}{(-)} \stackrel{13 14}{(-)}\,.
\end{eqnarray}
One can find both states in Table~\ref{Table III.}, the first $u_{R}$ as $u_{R8}$ in the eighth line
of this table, the second one as $u_{R7}$ in the seventh line of this table.

Below some useful relations follow.
From Eq.(\ref{snmb:gammatildegamma}) one has
\begin{eqnarray}
\label{stildestrans}
S^{ac}\stackrel{ab}{(k)}\stackrel{cd}{(k)} &=& -\frac{i}{2} \eta^{aa} \eta^{cc} 
\stackrel{ab}{[-k]}\stackrel{cd}{[-k]}\,,\,\quad\quad
\tilde{S}^{ac}\stackrel{ab}{(k)}\stackrel{cd}{(k)} = \frac{i}{2} \eta^{aa} \eta^{cc} 
\stackrel{ab}{[k]}\stackrel{cd}{[k]}\,,\,\nonumber\\
S^{ac}\stackrel{ab}{[k]}\stackrel{cd}{[k]} &=& \frac{i}{2}  
\stackrel{ab}{(-k)}\stackrel{cd}{(-k)}\,,\,\quad\quad
\tilde{S}^{ac}\stackrel{ab}{[k]}\stackrel{cd}{[k]} = -\frac{i}{2}  
\stackrel{ab}{(k)}\stackrel{cd}{(k)}\,,\,\nonumber\\
S^{ac}\stackrel{ab}{(k)}\stackrel{cd}{[k]}  &=& -\frac{i}{2} \eta^{aa}  
\stackrel{ab}{[-k]}\stackrel{cd}{(-k)}\,,\,\quad\quad
\tilde{S}^{ac}\stackrel{ab}{(k)}\stackrel{cd}{[k]} = -\frac{i}{2} \eta^{aa}  
\stackrel{ab}{[k]}\stackrel{cd}{(k)}\,,\,\nonumber\\
S^{ac}\stackrel{ab}{[k]}\stackrel{cd}{(k)} &=& \frac{i}{2} \eta^{cc}  
\stackrel{ab}{(-k)}\stackrel{cd}{[-k]}\,,\,\quad\quad
\tilde{S}^{ac}\stackrel{ab}{[k]}\stackrel{cd}{(k)} = \frac{i}{2} \eta^{cc}  
\stackrel{ab}{(k)}\stackrel{cd}{[k]}\,. 
\end{eqnarray}
We conclude from the above equation that $\tilde{S}^{ab}$ generate the 
equivalent representations with respect to $S^{ab}$ and opposite. 

We recognize in Eq.~(\ref{graphbinoms}) 
the demonstration of the nilpotent and the projector character of the Clifford algebra objects 
$\stackrel{ab}{(k)}$ and $\stackrel{ab}{[k]}$, respectively. 
\begin{eqnarray}
\stackrel{ab}{(k)}\stackrel{ab}{(k)}& =& 0\,, \quad \quad \stackrel{ab}{(k)}\stackrel{ab}{(-k)}
= \eta^{aa}  \stackrel{ab}{[k]}\,, \quad \stackrel{ab}{(-k)}\stackrel{ab}{(k)}=
\eta^{aa}   \stackrel{ab}{[-k]}\,,\quad
\stackrel{ab}{(-k)} \stackrel{ab}{(-k)} = 0\,, \nonumber\\
\stackrel{ab}{[k]}\stackrel{ab}{[k]}& =& \stackrel{ab}{[k]}\,, \quad \quad
\stackrel{ab}{[k]}\stackrel{ab}{[-k]}= 0\,, \;\;\quad \quad  \quad \stackrel{ab}{[-k]}\stackrel{ab}{[k]}=0\,,
 \;\;\quad \quad \quad \quad \stackrel{ab}{[-k]}\stackrel{ab}{[-k]} = \stackrel{ab}{[-k]}\,,
 \nonumber\\
\stackrel{ab}{(k)}\stackrel{ab}{[k]}& =& 0\,,\quad \quad \quad \stackrel{ab}{[k]}\stackrel{ab}{(k)}
=  \stackrel{ab}{(k)}\,, \quad \quad \quad \stackrel{ab}{(-k)}\stackrel{ab}{[k]}=
 \stackrel{ab}{(-k)}\,,\quad \quad \quad 
\stackrel{ab}{(-k)}\stackrel{ab}{[-k]} = 0\,,
\nonumber\\
\stackrel{ab}{(k)}\stackrel{ab}{[-k]}& =&  \stackrel{ab}{(k)}\,,
\quad \quad \stackrel{ab}{[k]}\stackrel{ab}{(-k)} =0,  \quad \quad 
\quad \stackrel{ab}{[-k]}\stackrel{ab}{(k)}= 0\,, \quad \quad \quad \quad
\stackrel{ab}{[-k]}\stackrel{ab}{(-k)} = \stackrel{ab}{(-k)}.
\label{graphbinoms}
\end{eqnarray}
%

%Defining
%%
%\begin{eqnarray}
%\stackrel{ab}{\tilde{(\pm i)}} = 
%\frac{1}{2} \, (\tilde{\gamma}^a \mp \tilde{\gamma}^b)\,, \quad
%\stackrel{ab}{\tilde{(\pm 1)}} = 
%\frac{1}{2} \, (\tilde{\gamma}^a \pm i\tilde{\gamma}^b)\,, 
%%\stackrel{ab}{\tilde{[\pm i]}} = \frac{1}{2} (1 \pm \tilde{\gamma}^a \tilde{\gamma}^b), \quad
%%\stackrel{ab}{\tilde{[\pm 1]}} = \frac{1}{2} (1 \pm i \tilde{\gamma}^a \tilde{\gamma}^b). \nonumber
%\label{deftildefun}
%\end{eqnarray}
%%
%one recognizes that
%%
%\begin{eqnarray}
%\stackrel{ab}{\tilde{( k)}} \, \stackrel{ab}{(k)}& =& 0\,, 
%\quad \;
%\stackrel{ab}{\tilde{(-k)}} \, \stackrel{ab}{(k)} = -i \eta^{aa}\,  \stackrel{ab}{[k]}\,,
%\quad\;
%\stackrel{ab}{\tilde{( k)}} \, \stackrel{ab}{[k]} = i\, \stackrel{ab}{(k)}\,,
%\quad\;
%\stackrel{ab}{\tilde{( k)}}\, \stackrel{ab}{[-k]} = 0\,.
%\label{graphbinomsfamilies}
%\end{eqnarray}
%%
%Recognizing that
%%
%\begin{eqnarray}
%\stackrel{ab}{(k)}^{\dagger}=\eta^{aa}\stackrel{ab}{(-k)}\,,\quad
%\stackrel{ab}{[k]}^{\dagger}= \stackrel{ab}{[k]}\,,
%\label{graphherstr}
%\end{eqnarray}
%%

Below some more useful relations~\cite{pikanorma} are presented: 
\begin{eqnarray}
\label{plusminus}
N^{\pm}_{+}         &=& N^{1}_{+} \pm i \,N^{2}_{+} = 
 - \stackrel{03}{(\mp i)} \stackrel{12}{(\pm )}\,, \quad N^{\pm}_{-}= N^{1}_{-} \pm i\,N^{2}_{-} = 
  \stackrel{03}{(\pm i)} \stackrel{12}{(\pm )}\,,\nonumber\\
\tilde{N}^{\pm}_{+} &=& - \stackrel{03}{\tilde{(\mp i)}} \stackrel{12}{\tilde{(\pm )}}\,, \quad 
\tilde{N}^{\pm}_{-}= %\tilde{N}^{1}_{-} \pm i\,\tilde{N}^{2}_{-} = 
  \stackrel{03} {\tilde{(\pm i)}} \stackrel{12} {\tilde{(\pm )}}\,,\nonumber\\ 
\tau^{1\pm}         &=& (\mp)\, \stackrel{56}{(\pm )} \stackrel{78}{(\mp )} \,, \quad   
\tau^{2\mp}=            (\mp)\, \stackrel{56}{(\mp )} \stackrel{78}{(\mp )} \,,\nonumber\\ 
\tilde{\tau}^{1\pm} &=& (\mp)\, \stackrel{56}{\tilde{(\pm )}} \stackrel{78}{\tilde{(\mp )}}\,,\quad   
\tilde{\tau}^{2\mp}= (\mp)\, \stackrel{56}{\tilde{(\mp )}} \stackrel{78}{\tilde{(\mp )}}\,.
\end{eqnarray}
%

%One Weyl representation of $SO(13+1)$  contains left handed weak charged and 
%the second $SU(2)$ chargeless coloured quarks and colourless leptons and right handed weak chargeless
%and the second $SU(2)$ charged quarks and leptons (electrons and neutrinos). It carries also the family
%quantum numbers, not mentioned in this table. The table is taken from Ref.~\cite{portoroz03}.
%
In Table~\ref{Table III.}~\cite{JMP2015} the eight families of the first member in  
Table~\ref{Table so13+1.} (member number  $1$) of the eight-plet of quarks and the $25^{th}$ 
member  in Table~\ref{Table so13+1.} of the eight-plet of leptons are presented as an example.
The eight families of the right handed $u_{1R}$ quark  are 
presented in the left column of  Table~\ref{Table III.}~\cite{JMP2015}. In the right column of the 
same table the equivalent eight-plet of the right handed neutrinos $\nu_{1R}$ are presented.
All the other members of any of the eight families of quarks or leptons follow  from any member 
of a particular family by the application of the  operators $N^{\pm}_{R,L}$ and  
$\tau^{(2,1)\pm}$, Eq.~(\ref{plusminus}) on this particular member.  

%\clearpage
 
%
 \begin{table}
 \begin{center}
 \begin{tabular}{|c|c|c|c|c|r r r r r|}
 \hline
 &&&&&$\tilde{\tau}^{13}$&$\tilde{\tau}^{23}$&$\tilde{N}_{L}^{3}$&$\tilde{N}_{R}^{3}$&$\tilde{\tau}^{4}$\\
 \hline
 $I$&$u^{c1}_{R\,1}$&
   $ \stackrel{03}{(+i)}\,\stackrel{12}{[+]}|\stackrel{56}{[+]}\,\stackrel{78}{(+)} ||
   \stackrel{9 \;10}{(+)}\;\;\stackrel{11\;12}{(-)}\;\;\stackrel{13\;14}{(-)}$ & 
   $\nu_{R\,1}$&
   $ \stackrel{03}{(+i)}\,\stackrel{12}{[+]}|\stackrel{56}{[+]}\,\stackrel{78}{(+)} ||
   \stackrel{9 \;10}{(+)}\;\;\stackrel{11\;12}{[+]}\;\;\stackrel{13\;14}{[+]}$ 
  &$-\frac{1}{2}$&$0$&$-\frac{1}{2}$&$0$&$\frac{1}{6}$ 
 \\
  $I$&$u^{c1}_{R\,2}$&
   $ \stackrel{03}{[+i]}\,\stackrel{12}{(+)}|\stackrel{56}{[+]}\,\stackrel{78}{(+)} ||
   \stackrel{9 \;10}{(+)}\;\;\stackrel{11\;12}{(-)}\;\;\stackrel{13\;14}{(-)}$ & 
   $\nu_{R\,2}$&
   $ \stackrel{03}{[+i]}\,\stackrel{12}{(+)}|\stackrel{56}{[+]}\,\stackrel{78}{(+)} ||
   \stackrel{9 \;10}{(+)}\;\;\stackrel{11\;12}{[+]}\;\;\stackrel{13\;14}{[+]}$ 
  &$-\frac{1}{2}$&$0$&$\frac{1}{2}$&$0$&$\frac{1}{6}$
 \\
  $I$&$u^{c1}_{R\,3}$&
   $ \stackrel{03}{(+i)}\,\stackrel{12}{[+]}|\stackrel{56}{(+)}\,\stackrel{78}{[+]} ||
   \stackrel{9 \;10}{(+)}\;\;\stackrel{11\;12}{(-)}\;\;\stackrel{13\;14}{(-)}$ & 
   $\nu_{R\,3}$&
   $ \stackrel{03}{(+i)}\,\stackrel{12}{[+]}|\stackrel{56}{(+)}\,\stackrel{78}{[+]} ||
   \stackrel{9 \;10}{(+)}\;\;\stackrel{11\;12}{[+]}\;\;\stackrel{13\;14}{[+]}$ 
  &$\frac{1}{2}$&$0$&$-\frac{1}{2}$&$0$&$\frac{1}{6}$
 \\
 $I$&$u^{c1}_{R\,4}$&
  $ \stackrel{03}{[+i]}\,\stackrel{12}{(+)}|\stackrel{56}{(+)}\,\stackrel{78}{[+]} ||
  \stackrel{9 \;10}{(+)}\;\;\stackrel{11\;12}{(-)}\;\;\stackrel{13\;14}{(-)}$ & 
  $\nu_{R\,4}$&
  $ \stackrel{03}{[+i]}\,\stackrel{12}{(+)}|\stackrel{56}{(+)}\,\stackrel{78}{[+]} ||
  \stackrel{9 \;10}{(+)}\;\;\stackrel{11\;12}{[+]}\;\;\stackrel{13\;14}{[+]}$ 
  &$\frac{1}{2}$&$0$&$\frac{1}{2}$&$0$&$\frac{1}{6}$
  \\
  \hline
  $II$& $u^{c1}_{R\,5}$&
        $ \stackrel{03}{[+i]}\,\stackrel{12}{[+]}|\stackrel{56}{[+]}\,\stackrel{78}{[+]}||
        \stackrel{9 \;10}{(+)}\;\;\stackrel{11\;12}{(-)}\;\;\stackrel{13\;14}{(-)}$ & 
        $\nu_{R\,5}$&
        $ \stackrel{03}{[+i]}\,\stackrel{12}{[+]}|\stackrel{56}{[+]}\,\stackrel{78}{[+]}|| 
        \stackrel{9 \;10}{(+)}\;\;\stackrel{11\;12}{[+]}\;\;\stackrel{13\;14}{[+]}$ 
        &$0$&$-\frac{1}{2}$&$0$&$-\frac{1}{2}$&$\frac{1}{6}$
 \\ 
  $II$& $u^{c1}_{R\,6}$&
      $ \stackrel{03}{(+i)}\,\stackrel{12}{(+)}|\stackrel{56}{[+]}\,\stackrel{78}{[+]}||
      \stackrel{9 \;10}{(+)}\;\;\stackrel{11\;12}{(-)}\;\;\stackrel{13\;14}{(-)}$ & 
      $\nu_{R\,6}$&
      $ \stackrel{03}{(+i)}\,\stackrel{12}{(+)}|\stackrel{56}{[+]}\,\stackrel{78}{[+]}|| 
      \stackrel{9 \;10}{(+)}\;\;\stackrel{11\;12}{[+]}\;\;\stackrel{13\;14}{[+]}$ 
      &$0$&$-\frac{1}{2}$&$0$&$\frac{1}{2}$&$\frac{1}{6}$
 \\ 
 $II$& $u^{c1}_{R\,7}$&
 $ \stackrel{03}{[+i]}\,\stackrel{12}{[+]}|\stackrel{56}{(+)}\,\stackrel{78}{(+)}||
 \stackrel{9 \;10}{(+)}\;\;\stackrel{11\;12}{(-)}\;\;\stackrel{13\;14}{(-)}$ & 
      $\nu_{R\,7}$&
      $ \stackrel{03}{[+i]}\,\stackrel{12}{[+]}|\stackrel{56}{(+)}\,\stackrel{78}{(+)}|| 
      \stackrel{9 \;10}{(+)}\;\;\stackrel{11\;12}{[+]}\;\;\stackrel{13\;14}{[+]}$ 
    &$0$&$\frac{1}{2}$&$0$&$-\frac{1}{2}$&$\frac{1}{6}$
  \\
   $II$& $u^{c1}_{R\,8}$&
    $ \stackrel{03}{(+i)}\,\stackrel{12}{(+)}|\stackrel{56}{(+)}\,\stackrel{78}{(+)}||
    \stackrel{9 \;10}{(+)}\;\;\stackrel{11\;12}{(-)}\;\;\stackrel{13\;14}{(-)}$ & 
    $\nu_{R\,8}$&
    $ \stackrel{03}{(+i)}\,\stackrel{12}{(+)}|\stackrel{56}{(+)}\,\stackrel{78}{(+)}|| 
    \stackrel{9 \;10}{(+)}\;\;\stackrel{11\;12}{[+]}\;\;\stackrel{13\;14}{[+]}$ 
    &$0$&$\frac{1}{2}$&$0$&$\frac{1}{2}$&$\frac{1}{6}$
 \\ 
 \hline 
 \end{tabular}
 \end{center}
\caption{\label{Table III.} 
Eight families of the right handed $u^{c1}_{R}$ (\ref{Table so13+1.}) 
quark with spin $\frac{1}{2}$, the colour charge $(\tau^{33}=1/2$, $\tau^{38}=1/(2\sqrt{3})$
(the definition of the operators is presented in Eqs.~(\ref{so42},\ref{so64}), a few examples
 how to calculate the application of these operators on the states can be found in 
Subsect.~\ref{explicit}.
The definition of the operators, expressible with $\tilde{S}^{ab}$ is: $\vec{\tilde{N}}_{L,R}$
 $= \frac{1}{2} (\tilde{S}^{23}\pm i \tilde{S}^{01}$, $
\tilde{S}^{31}\pm i \tilde{S}^{02}$, $\tilde{S}^{12}\pm i \tilde{S}^{03})$,
$\vec{\tilde{\tau}}^{1}=$ $\frac{1}{2} (\tilde{S}^{58}-  \tilde{S}^{67}$, $\tilde{S}^{57} + 
 \tilde{S}^{68}$, $\tilde{S}^{56}-  \tilde{S}^{78})$, 
 $\vec{\tilde{\tau}}^{2}$ $=\frac{1}{2} (\tilde{S}^{58}+  \tilde{S}^{67}$, $\tilde{S}^{57} - 
 \tilde{S}^{68}$, $\tilde{S}^{56}+  \tilde{S}^{78}) $ and $\tilde{\tau}^{4} $ $= -\frac{1}{3} 
(\tilde{S}^{9\;10} + \tilde{S}^{11\;12} + \tilde{S}^{13\;14})$), 
and of  the colourless right handed neutrino $\nu_{R}$ of spin $\frac{1}{2}$ %(\ref{Table II.})  
are presented in the  left and in the right column, respectively.
They belong to two groups of four families, one ($I$) is a doublet with respect to 
($\vec{\tilde{N}}_{L}$ and  $\vec{\tilde{\tau}}^{1}$) and  a singlet with respect to 
($\vec{\tilde{N}}_{R}$ and  $\vec{\tilde{\tau}}^{2}$), the other ($II$) is a singlet with respect to 
($\vec{\tilde{N}}_{L}$ and  $\vec{\tilde{\tau}}^{1}$) and  a doublet with with respect to 
($\vec{\tilde{N}}_{R}$ and  $\vec{\tilde{\tau}}^{2}$).
All the families follow from the starting one by the application of the operators 
($\tilde{N}^{\pm}_{R,L}$, $\tilde{\tau}^{(2,1)\pm}$), Eq.~(\ref{plusminus}).  The generators 
($N^{\pm}_{R,L} $, $\tau^{(2,1)\pm}$) (Eq.~(\ref{plusminus}))
transform $u_{Ri}, i=(1,\cdots,8),$ to all the members of the same colour of the family.  
The same generators transform equivalently the right handed   neutrino $\nu_{Ri},  i=(1,\cdots,8)$,
 to all the colourless members of the  $i^{th}$  family.
}
 \end{table}
The eight-plets separate into two group of four families: One group  contains  doublets with respect 
to $\vec{\tilde{N}}_{R}$ and  $\vec{\tilde{\tau}}^{2}$, these families are singlets with respect to 
$\vec{\tilde{N}}_{L}$ and  $\vec{\tilde{\tau}}^{1}$. Another group of families contains  doublets 
with respect to  $\vec{\tilde{N}}_{L}$ and  $\vec{\tilde{\tau}}^{1}$, these families are singlets 
with respect to  $\vec{\tilde{N}}_{R}$ and  $\vec{\tilde{\tau}}^{2}$. 

The scalar fields which are the gauge scalars  of  $\vec{\tilde{N}}_{R}$ and  $\vec{\tilde{\tau}}^{2}$ 
couple only to the four families  which are doublets with respect to these two groups. 
The scalar fields which are the gauge scalars  of  $\vec{\tilde{N}}_{L}$ and  $\vec{\tilde{\tau}}^{1}$ 
couple only to the four families  which are doublets with respect to these last two groups. 

After the electroweak phase transition, caused by the scalar fields with the space index $(7,8)$,
the two groups of four families become massive. The lowest of the two groups of four families 
contains the observed three, while the fourth remains to be measured. The lowest of the upper 
four families is the candidate  for the dark matter~\cite{IARD2016}.


\begin{thebibliography}{99}
\bibitem{NPLB} N.S. Manko\v c Bor\v stnik, "Do we have the explanation for
the Higgs and Yukawa couplings of the {\em standard model}", [arXiv:1212.3184,
arXiv:1011.5765].
\end{thebibliography}
\end{document}